\input{psfig}
\documentclass[]{aa}
\usepackage{graphicx}
\usepackage{color,natbib}
\usepackage{deluxetable}
\begin{document}
\title{ The $\beta$ Pictoris association low-mass members: membership assessment, rotation period distribution, and dependence on multiplicity\thanks{Tables 2-3 are only available in electronic form
at the CDS via anonymous ftp to cdsarc.u-strasbg.fr (130.79.128.5)
or via http://cdsweb.u-strasbg.fr/cgi-bin/qcat?J/A+A/.}}
\author{S.\,Messina\inst{1},  A.C.\,Lanzafame\inst{2,1},  L.\,Malo\inst{3}, S.\,Desidera\inst{4}, A.\,Buccino\inst{5,6},  L.\,Zhang \inst{7}, S.\,Artemenko\inst{8} M.\,Millward\inst{9}, F.-J.\,Hambsch\inst{10,11}}
\offprints{Sergio Messina}
\institute{INAF-Catania Astrophysical Observatory, via S.Sofia, 78 I-95123 Catania, Italy \\
\email{sergio.messina@oact.inaf.it}
\and
Universit\'a di Catania, Dipartimento di Fisica e Astronomia, Sezione Astrofisica, via S. Sofia 78, I-95123 Catania, Italy\\
\email{a.lanzafame@unict.it}
\and  
Canada-France-Hawaii Telescope, 65-1238 Mamalahoa Hwy, Kamuela, HI 96743, USA\\
\email{malo@cfht.hawaii.edu}
\and
INAF-Osservatorio Astronomico di Padova, Vicolo dell'Osservatorio 5, I-35122 Padova, Italy\\
\email{silvano.desidera@oapd.inaf.it}
\and
Instituto de Astronom\'ia y F\'isica del Espacio (IAFE-CONICET), Buenos Aires, Argentina\\
\email{abuccino@iafe.uba.ar}
\and
Departamento de F\'isica, FCEN-Universidad de Buenos Aires, Argentina
\and
Department of Physics, College of Science, Guizhou University, Guiyang 550025, P.R. China\\
\email{liy\_zhang@hotmail.com}
\and
Research Institute Crimean Astrophysical Observatory, 298409, Nauchny, Crimea\\
\email{svetaartemenko@rambler.ru}
\and   
York Creek Observatory, Georgetown, Tasmania, Australia \\
\email{mervyn.millward@yorkcreek.net}
\and
Remote Observatory Atacama Desert (ROAD), Vereniging Voor Sterrenkunde (VVS), Oude Bleken 12, B-2400 Mol, Belgium\\
\email{hambsch@telenet.be}
\and
American Association of Variable Star Observers (AAVSO), Cambridge, MA, USA}
\date{}
\titlerunning{Membership and rotation period distribution among $\beta$ Pic members}
\authorrunning{S.\,Messina et al.}
\abstract {Low-mass members of young loose stellar associations and open clusters exhibit a wide spread of rotation periods. Such a spread originates from distributions of masses and initial rotation periods. However,   multiplicity can also play a significant role.}
{We want to investigate the role played by   physical companions in  multiple systems in shortening the primordial disc lifetime, anticipating the rotation spin up with respect to single stars. } {We have compiled the most extensive to date list of low-mass bona fide and candidate members of the young 25-Myr $\beta$ Pictoris association. We have measured from our own photometric time series or from archival time series the rotation periods of about all members. In a few cases the rotation periods were retrieved from the literature. We used updated UVWXYZ components to assess the membership of the whole stellar sample. Thanks to the known basic properties of most members we built the rotation period distribution distinguishing between bona fide members and candidate members and according to their multiplicity status. } { We found that single stars and components of multiple systems in wide orbits ($>$80\,AU) have rotation periods that exhibit a well defined sequence arising from  mass distribution  with some level of spread likely arising from  initial rotation period distribution. All components of multiple systems in close orbits ($<$80\,AU) have rotation periods significantly shorter than their equal-mass single counterparts.  For these close components of multiple systems a linear dependence of rotation rate on separation is only barely detected. 
 A comparison with the younger 13 Myr $h$ Per cluster and with the older 40-Myr open clusters/stellar associations NGC2547, IC\,2391, Argus, and IC\,2602 and the 130-Myr Pleiades shows that whereas the evolution of  F-G stars is well reproduced by angular momentum evolution models, this is not the case for the slow K and early-M stars.  Finally, we found that the amplitude of their light curves is correlated neither with rotation nor with mass.} {Once single stars and wide components of multiple systems are separated from close components of multiple systems, the rotation period distributions exhibit a well defined dependence on mass that allows to make a meaningful comparison with similar distributions of either younger or older associations/clusters. Such cleaned distributions allow to use the stellar rotation period as age indicator, meaningfully for F and G type stars. }
\maketitle
\rm

\section{Introduction}
	$\beta$ Pictoris is a nearby young loose stellar association. Its members have an average distance from the Sun of about 40$\pm$17\,pc   and an age of about 25$\pm$3\,Myr (\citealt{Messina16a}; hereafter Paper I). Youth and proximity make this association a special  benchmark in stellar astrophysics studies. In fact, the  young age secures the presence of interesting circumstellar environments in many members, where discs and planetary systems can be discovered. The proximity allows to spatially resolve them giving effective possibility to study disc morphology and the planetary system's architecture. The A3V star $\beta$ Pictoris, from which the association takes the name, is one such example (see, e.g., \citealt{Chauvin12}). Youth, vicinity, and brightness of its members explain why this association has been included in many studies to search for very low-mass stellar and planetary companions, as well as to accurately measure element abundances, magnetic activity, and kinematics. Among many studies, we mention those aimed at searching for planetary companions and discs like the \it SEEDS \rm project (Strategic Exploration of Exoplanets and Disks with Subaru; \citealt{Brandt14}),  \it The Gemini/NICI planet-finding campaign \rm (\citealt{Biller13}), \it SPHERE \rm (Spectro-Polarimetric High-contrast Exoplanet REsearch, \citealt{Beuzit08}), and \it NaCo Large Program \rm (\citealt{Desidera15}); those aimed at searching and characterizing new members, like \it SACY \rm  project (Search for Associations Containing Young stars; \citealt{Torres06}, \citeyear{Torres08}; \citealt{daSilva09}; \citealt{Elliott14}), \it The solar Neighborhood \rm  investigation (\citealt{Riedel14}), the \it BANYAN \rm project (Bayesian Analysis for Nearby Young Associations; \citealt{Malo13}, \citeyear{Malo14a}, \citeyear{Malo14b}), and many other membership investigations (see, e.g., \citealt{Lepine09}; \citealt{Kiss11}; \citealt{Schlieder10}, \citeyear{Schlieder12}; \citealt{Shkolnik12}; \citealt{Malo13}, \citeyear{Malo14a}, \citeyear{Malo14b}) resulting in a significantly increased number, by about a factor 3, of confirmed and new candidate members, with respect to the association members detected in discovery studies  (e.g., \citealt{Zuckerman01}).\\
The first comprehensive rotational investigation of the low-mass (spectral types F to M) members of  the $\beta$ Pictoris association was carried out by Messina et al. (\citeyear{Messina10}, \citeyear{Messina11}). They measured the rotation periods of 33 on a list of 38 among confirmed and candidate members compiled from  Zuckerman \& Song (2004) and Torres et al. (2006, 2008).\\
The rotational properties of the $\beta$ Pictoris association represent a key information for a number of  studies concerning, e.g., the pre-main-sequence (PMS) angular momentum evolution of low-mass stars (see \citealt{Spada11}; \citealt{Gallet13},  \citeyear{gallet15}), the effect of rotation on Lithium depletion at young ages (\citealt{Pallavicini93}; \citealt{Bouvier16}; \citealt{Messina16a}),  the impact of photoevaporation and binary encounters   on the primordial disc life time (\citealt{Olczak10}; \citealt{Throop08}) and the time scale  of the star-disc locking phase, which can all be probed by means of the star's rotation (see, e.g., \citealt{Messina14}, \citeyear{Messina15a}, \citeyear{Messina15b}), as well as the implication for the formation of planets around binaries (see, e.g., \citealt{Alexander12}).\\
Considering the importance of  the $\beta$ Pictoris association for these studies, the increased number of newly discovered members, and the fact that the basic properties of many members have been time by time better established (thanks to their brightness),  we realized that the time was ripe for carrying out a new rotational study on this enlarged sample. The results of this extensive study were presented in the catalog of photometric rotational periods by Messina et al. (\citeyear{Messina16b}; hereafter Paper II) containing the photometric rotational periods of 112 low-mass members and candidate members of the $\beta$ Pictoris association. These rotation periods were used to explore the rotation-Lithium connection and to obtain an improved age estimate of the $\beta$ Pictoris association using the Lithium Depletion Boundary method (Paper I). In the present study (Paper III), we aim at exploiting this  catalogue of rotation periods  to investigate the distribution of rotation periods versus mass and the role played by multiplicity, which is known for most members, in determining the wide spread of rotation periods observed in this and other young loose associations.\\
In Sect.\,2, we present the up-to-date and most complete sample of members. In Sect.\,3, we discuss on the basic properties, color and rotation period, that are used in our analysis. In Sect.\,4, we present the results of our period search. In Sect.\,5, we assess the membership of the whole sample using updated space and velocity components. In Sect.\,6, we discuss the rotation period distribution and present new results on the impact of multiplicity on the rotation evolution. In Sect.\,7, we make a comparison of the rotation period distribution with those of  younger and of older open clusters and associations. Dependence of photospheric activity, as measured from light curve amplitude, on rotation and mass is discussed in Sect.\,8. In Sect.\,9, we give our conclusions.
\begin{figure}
\begin{minipage}{10cm}
\includegraphics[scale = 0.5, trim = 0 0 0 0, clip, angle=0]{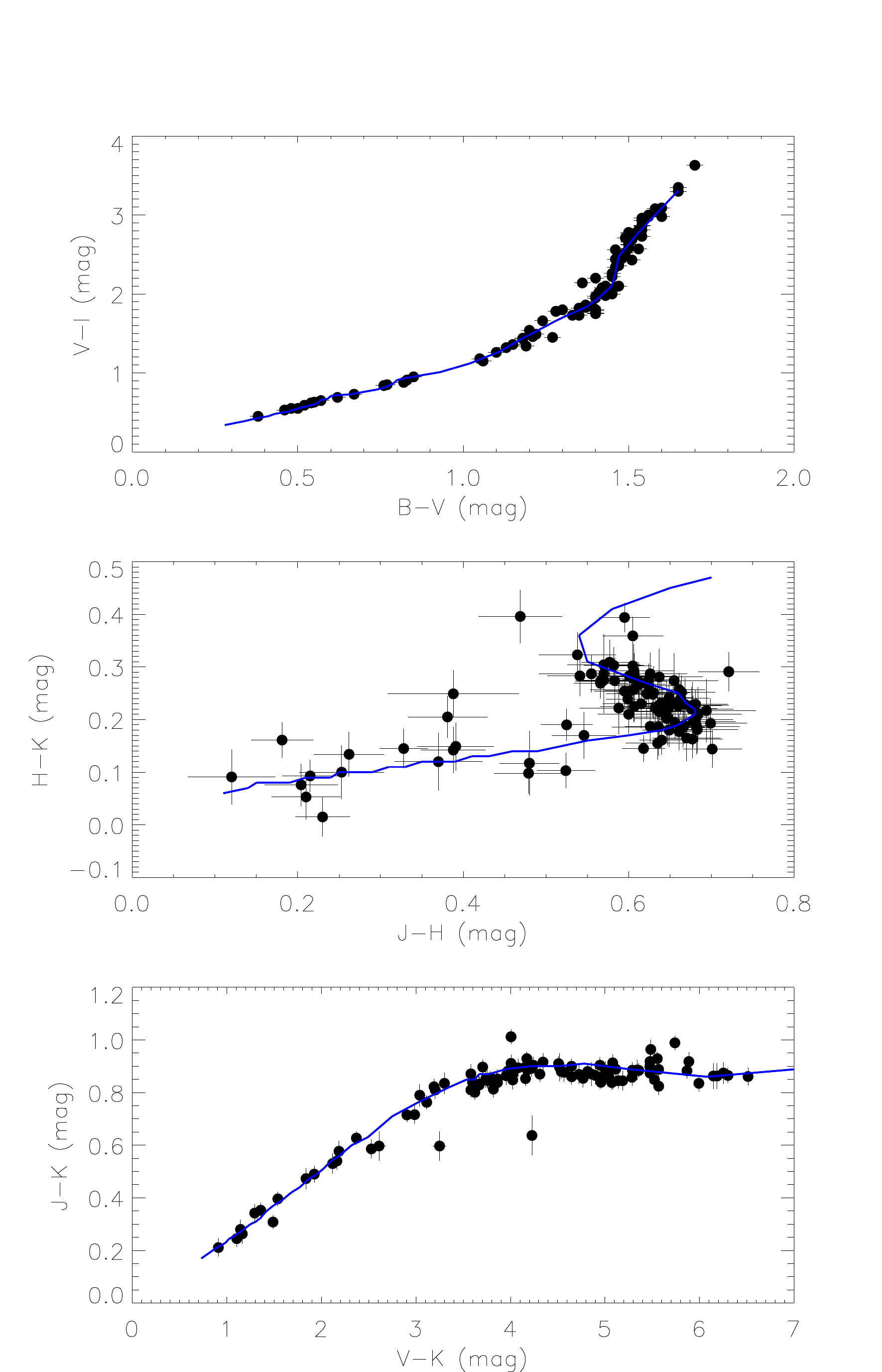}
\end{minipage}
\caption{\label{color-color} Color-color plots for the $\beta$ Pictoris members and candidate members
with overplotted polynomial fits to the corresponding colors  taken from Table\,6 of \citet{Pecaut13}.}
\end{figure}

\section{Sample description}
The present study is based on the catalog
of photometric rotational periods of low-mass members and candidate members of the $\beta$ Pictoris association presented  in Paper II. Briefly, we have carried out an extensive search in the literature to retrieve all  members of the $\beta$ Pictoris association.
We compiled a list of 117 among bona fide  and candidate members, with spectral types later than about F3V, from the following major studies: Torres et al. (\citeyear{Torres06}, \citeyear{Torres08}), \citet{Lepine09}, \citet{Kiss11}, Schlieder et al. (\citeyear{Schlieder10}, \citeyear{Schlieder12}), \citet{Shkolnik12}, Malo et al. (\citeyear{Malo13}, \citeyear{Malo14a}, \citeyear{Malo14b}), and other studies detailed for each member in Paper II.
 Stars of earlier spectral types were excluded from our sample since the photometric rotation period to be measured requires the presence of a detectable level of magnetic activity (more specifically, of light rotational modulation by surface temperature inhomogeneities with amplitude of several millimag at least). This circumstance generally occurs  in late spectral type stars that have an external convection zone, which allows for the production of magnetic fields and is subjected to magnetic braking.\\ 
We measured the rotation periods of 112 out of 117 stars either from our own photometric monitoring or from photometric time series in public archives, or we retrieved these periods from the literature.\\
Information on individual stars either from our own analysis or from the literature and references can be retrieved in Paper II.
Information on the membership is not homogeneous for all the targets  either for the number of studies or for the 
methods. For example,  we found more than four membership studies for 52 targets, whereas only one membership study for 15 targets. For this reason, in Sect. 4, we present the results of our membership study based on updated space and velocity components and Lithium equivalent width (EW).\\
 The single/binary nature of our targets is based on the available RV measurements and direct imaging studies, which are referenced for each target in  Appendix A of Paper II. We note that not all stars with constant RV have been observed with high-contrast direct imaging. Therefore, for these stars, despite the RV constancy, we cannot rule out the presence of a wide orbit companion. However, even if this is the case, their rotational properties are indistinguishable from those of known wide binaries (see Sect.\,5). Targets with not determined either single or binary nature from RV studies are flagged with a symbol '?' in the last column of Table \ref{tab_period}. 
The complete target's list is reported in Table 1.\\
\section{Target properties}
\subsection{Colors}
Our aim is to investigate the distribution of the rotation periods versus stellar mass and the impact of multiplicity on the observed rotation period spread. The stellar mass for the majority of our target stars has to be derived from a comparison with evolutionary mass tracks at the age of the $\beta$ Pictoris association. The derived masses, especially for later spectral type stars, significantly depend on the adopted model, with models including effects of magnetic fields giving results more congruent with other age dating methods with respect to non magnetic models (see, \citealt{Messina16c} for a detailed discussion).
The associated uncertainty on the mass value derives from the uncertainties on distance, apparent magnitude, and effective temperature.
In most cases, effective temperatures, which are inferred from spectral typesÊ especially for the mid- to late-M stars, have uncertainties not smaller than $\pm$100\,K. 
For this reason, we have investigated which color index is the best stellar mass proxy.\\
In our sample, B$-$V and V$-$I are available for 60 targets; 41 targets have V$-$I only; 6 targets  have B$-$V only. All these colors are listed in Table\,1 of Paper II and were compiled from different sources in the literature. The remaining 10 targets have both B$-$V and V$-$I colors unknown. 
Since the color is a basic parameter in the following analysis, we had to recover the missing values.
In the top panel of Fig.\,\ref{color-color}, we plot the observed V$-$I versus B$-$V colors for the program stars.
We overplot (blue solid line) a polynomial fit to the intrinsic V$-$I versus B$-$V  colors 
listed by \citet{Pecaut13} for the 5--30 Myr old stars. 
The agreement is good with an average scatter  of 0.05\,mag of our colors from the polynomial relation. The agreement mainly arises from the fact that  there are a number of $\beta$ Pictoris members in our sample that were used by Pecaut \& Mamajek  to infer their tabulated colors for young stars.  
 We used this relation to derive the colors from the measured V$-$I and B$-$V colors, respectively,
 for the mentioned targets missing either B$-$V or V$-$I, and their associated uncertainty is 0.05\,mag.
 For the remaining 10 targets 
with no colors, we inferred them from the spectral type using again the Pecaut \& Mamajek color versus Spectral Type  relations,  with an associated uncertainty of 0.07\,mag.   For instance, the distances of our targets  have an average value of about 40\,pc, therefore, the interstellar reddening can be considered negligible and we did not apply any color correction in our analysis.\\
  We note that whereas the B$-$V color index of our targets spans a range $\Delta$$\sim$1.35 of magnitudes, the V$-$I color index  spans a much larger $\Delta$$\sim$3.3 magnitude range, then the latter color index is  better suited to represent stars of different masses.
 On the other hand, in addition to the limit arising from the use of derived V$-$I colors for about $\sim$70\% of the sample,  the two colors come from different works for the majority of stars and they were measured at different epochs. Due to magnetic activity,  colors can vary in time up to several hundredths of magnitude\footnote{The series of papers on the multiband photometric monitoring of active stars by Cutispoto et al. (e.g., \citeyear{Cutispoto03}, and references therein), provide an exhaustive example.} in such a young association. Therefore, the measured colors (and those derived) are not as homogeneous as we would like.\\
 \begin{figure*}
\begin{minipage}{10cm}
\includegraphics[scale = 0.7, trim = 0 0 0 80, clip, angle=90]{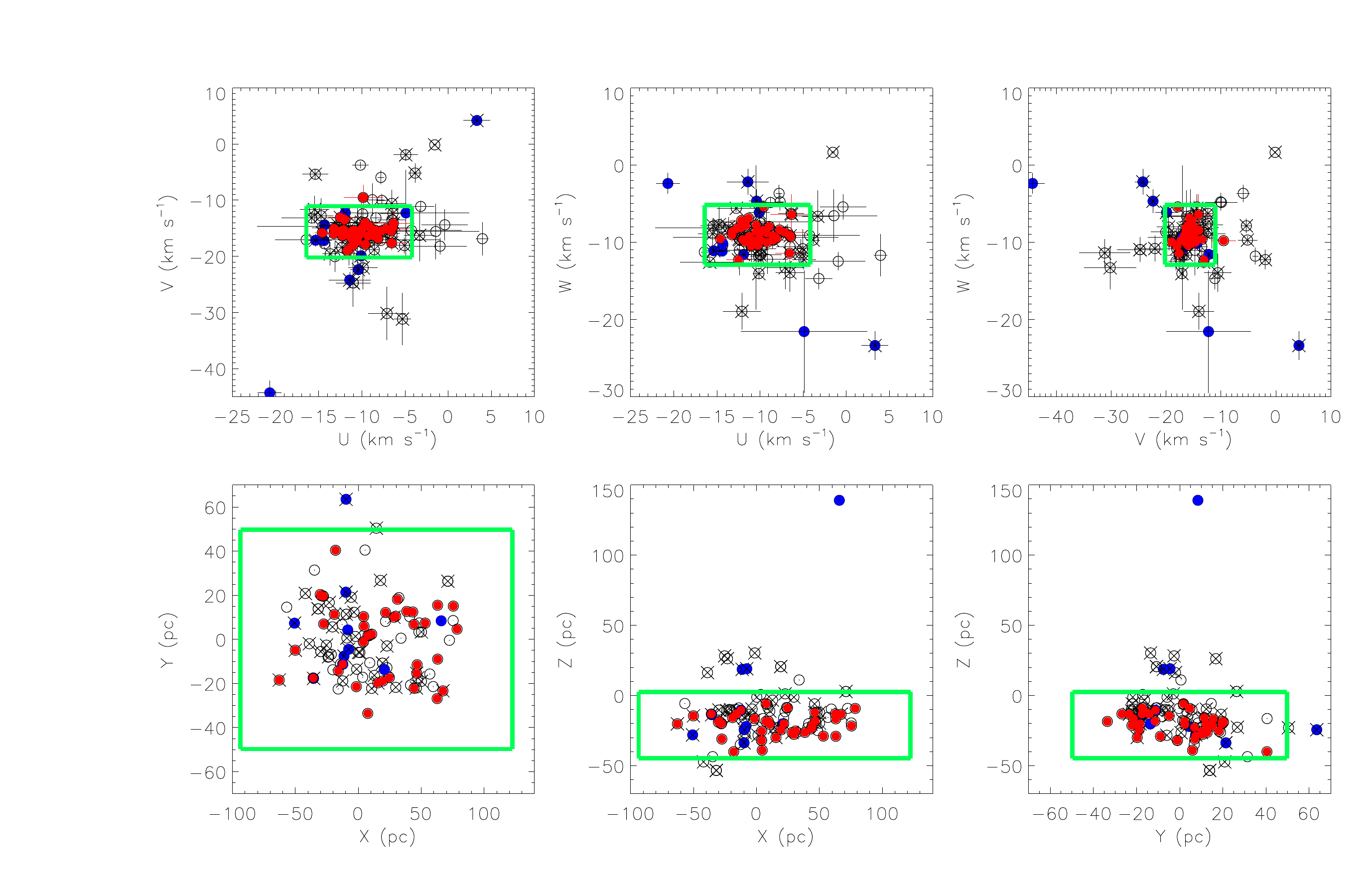}
\end{minipage}
\caption{\label{UVW} Distribution of UVW velocity (top panels) and XYZ space (bottom panels) Galactic components of members and candidate members of the $\beta$ Pictoris association. Red bullets represent the core sample,  open bullets the candidate members, blue bullets the rejected members, crossed symbols are stars with no Li measurement.  The green rectangular boxes identify the plane region within 3$\sigma$ from the average values (see text).}
\end{figure*}

 In the middle panel of Fig.\,\ref{color-color}, we plot the H$-$K$_s$ versus J$-$H colors of our targets measured by the 2MASS project (\citealt{Cutri03}). We overplot (blue solid line) a polynomial fit to the intrinsic H$-$K$_s$ versus J$-$H  colors 
listed by \citet{Pecaut13} for the 5--30 Myr old stars. We note that a few stars deviate significantly from this relation. They are close binaries with components of different spectral types that were unresolved by 2MASS. We note that the 2MASS color indices of our targets span a range of magnitudes not larger than $\Delta$$\sim$0.6, which is too small for our purposes, and, more importantly, the relation is not univocal.\\
 Finally, in the bottom panel of Fig.\,\ref{color-color}, we plot the J$-$K$_s$ versus V$-$K$_s$ colors of our targets measured by 2MASS, whereas the V magnitude is the one listed in Table\,1 of Paper II and taken (for 98 out of 117 stars)  as the brightest (and presumably unspotted) magnitude in the ASAS (All Sky Automated Survey; \citealt{Pojmanski97}) timeseries or as the brightest magnitude reported in literature (for the remaining 19 stars). Again, we note a few stars deviating significantly from the average trend. We find that the V$-$K$_s$ color index has a magnitude range of $\Delta$$\sim$7 and is the best suited to investigate the color-period distribution. The use of the V$-$K$_s$ color allows us to deal with an average  uncertainty from $\sim$2\% for K0V stars to less than $\sim$0.5\% for late-M stars.\\
 
 For instance, we note that a comparison with the polynomial fit from \citet{Pecaut13} shows that our targets that belong to multiple systems and are unresolved in the 2MASS photometry (separation $\rho \le 6^{\prime\prime}$ between the components) have V$-$K$_s$ colors redder on average by 0.03 mag with respect to resolved targets.\\

\subsection{Rotation period}
The other fundamental stellar property in our investigation is the rotation period.  To measure the photometric rotation periods of our targets, we used archive data, we made use of periods from the literature, and carried out our own multi-observatory observations.  A detailed description of the instruments, log of the observations, and information on data reduction and analysis, and the results of the period search are presented in Paper II.
\\
Briefly, 
in our sample, 52 stars have photometric time series in one or more of the following public archives: ASAS (All Sky Automated Survey; \citealt{Pojmanski97}), SuperWASP (Wide Angle Search for Planets; \citealt{Butters10}), Integral/OMC (\citealt{Domingo10}), Hipparcos (ESA 1997), NSVS (Northern Sky Variability Survey; \citealt{Wozniak04}), MEarth (\citealt{Berta12}), and CSS (Catalina Sky Survey; \citealt{Drake09}). We have retrieved and analysed  all the available time series for the period search. \\
Another  20 stars in our sample had no archive data and, thus, they were photometrically monitored by us for the first time.
We also observed another 15 stars that, although present in one of the mentioned archives, were either in close binary systems with
unresolved components or the archive data did not allow a period determination. We obtained either photometric time series for the resolved components or photometric  series suitable for a successful period measurement. For the remaining 30 stars we adopted the rotation periods available in the literature. The results are summarized in Table 2 of Paper II.\\
To search for the stellar rotation periods of our targets we have followed an approach similar to that used in Messina et al. (\citeyear{Messina10}, \citeyear{Messina11}). We refer the reader to those papers and to Paper II for a detailed description of the methods.\\
As a result of our photometric analysis, we obtained  the rotation period of 112 out of 117 target stars. Specifically,
we measured for the first time the rotation period of 56 stars. For another 27 stars, we confirmed the values reported in the literature with our analysis of
new or archived data. For 29 stars we adopted the literature values. For the remaining 5 stars, our periodogram analysis did not provide the rotation period.
\section{Membership assessment}
For a meaningful investigation of the rotation period distribution and dependences on multiplicity, we first carried out a membership assessment of all 117 stars in our sample  by comparing their Galactic velocity (UVW)\footnote{U positive towards the Galactic center, V positive in the direction of the Galactic rotation, and W positive in the direction of the Galactic north pole.} relative to the Sun and space (XYZ) components  with respect to the association average values. The proper motions, radial velocities, and distances used to derive UVW and their uncertainties, and XYZ are listed in Table 2 together with their references. We generally found more measurements of RV for each star in the literature and  measured a weighted average and its standard deviation   for our purposes. Individual RV measurements and relative references are listed in Table 3.\\
To measure the average values of the Galactic components, we selected an initial sub-sample consisting of stars that
were already known as bona fide members of the association and, more precisely, that were investigated in several earlier studies (up to eight for a few; see Paper II) that all agreed to assign the membership to the $\beta$ Pictoris association. \\
Among these stars, we subsequently selected only single and wide-orbit components of multiple systems to minimize the effect on the derived Galactic components of RV variation arising from orbital motion.  In such a way, we were left with 41 stars that represent our 'core' sample.\\
We use this core sample to compute the average  $\overline{U}$,  $\overline{V}$, and  $\overline{W}$ velocity components and their  standard deviations  $\sigma_U$,  $\sigma_V$, and  $\sigma_W$, and the average  $\overline{X}$,  $\overline{Y}$, and  $\overline{Z}$ space components  and their standard deviations. 
After computing average values and  standard deviations, we found six stars of the core sample that  significantly deviated ($>$ 3$\sigma$) from the other core members in two of three planes ([U,V], [U,W],[V,W]): four components of wide binaries (2MASS\,J02014677+0117161, RBS\,269, 2MASS\,J04435686+3723033, TYC\,6872\,1011\,1), and two single stars (2MASS\,J02175601+1225266, 2MASS\,J16430128-1754274 with very large uncertainties in their velocity components). 
These stars were excluded from the core sample and  new average values and 
 standard deviations  were recomputed as reported in the following:
\begin{equation}
 \overline{U} (km\,s^{-1}) = -10.27\pm1.68\\
      \end{equation}
      \begin{equation}
 \overline{V} (km\,s^{-1}) = -15.80\pm0.90\\
      \end{equation}
      \begin{equation}
 \overline{W} (km\,s^{-1}) =  -8.77\pm1.20\\
       \end{equation}
       \begin{equation}
   \overline{X} (pc)  =  18\pm32\\
       \end{equation}
       \begin{equation}
   \overline{ Y } (pc)   =  1\pm16\\
     \end{equation}
     \begin{equation}
    \overline{ Z } (pc)    =  -20\pm7\\
     \end{equation}
     In Fig.\,\ref{UVW}, we show the 6D kinematic distributions of all 117 stars in our sample. Red bullets represent our core members and the rectangular boxes identify the plane region within 3$\sigma$ from the average values. The values we derived are in agreement within the uncertainties with the values of \citet{Torres08}.\\
In addition to the kinematics, we used also the Li EW, whenever available (see Paper I for a list of targets with measured Li EW), as  a strong constrain  to asses the membership. Those stars in our sample whose UVW and XYZ differ by less than 3$\sigma$ from the average values of the core sample  but whose Li EW significantly deviates ($>$3$\sigma$) from the linear fits to the distribution exhibited by core members (see Fig.\,2 in Paper I)  are considered non members of the association.\\ 
The results of our membership assessment are summarized in Table 4.
As result, in our sample we have 80 bona fide members (flagged with 'Y'), of which 35 constituting the core sample (flagged with 'Core'),  that fully satisfy our criteria for membership, and the above mentioned six bona fide members as reported in the literature, but excluded from our core sample (flagged with 'Core\_e'). 
In the present study, we classify as candidate members those stars that have from one to three among space and velocity components deviating more than 3$\sigma$ from the average of the $\beta$ Pictoris association. Whereas, we classify as non members those stars with more than three among space and velocity components deviating more than 3$\sigma$ from the average. Accordingly, in our sample we have 22 candidate members (flagged with 'C') and 15 non members (flagged with 'NO').  We note that in the following the adjectives 'bona fide' and 'candidate' only refer to the membership status and not to the single/multiple nature of the targets. 
\begin{figure*}
\begin{minipage}{18cm}
\includegraphics[scale = 0.5, trim = 0 0 0 0, clip, angle=90]{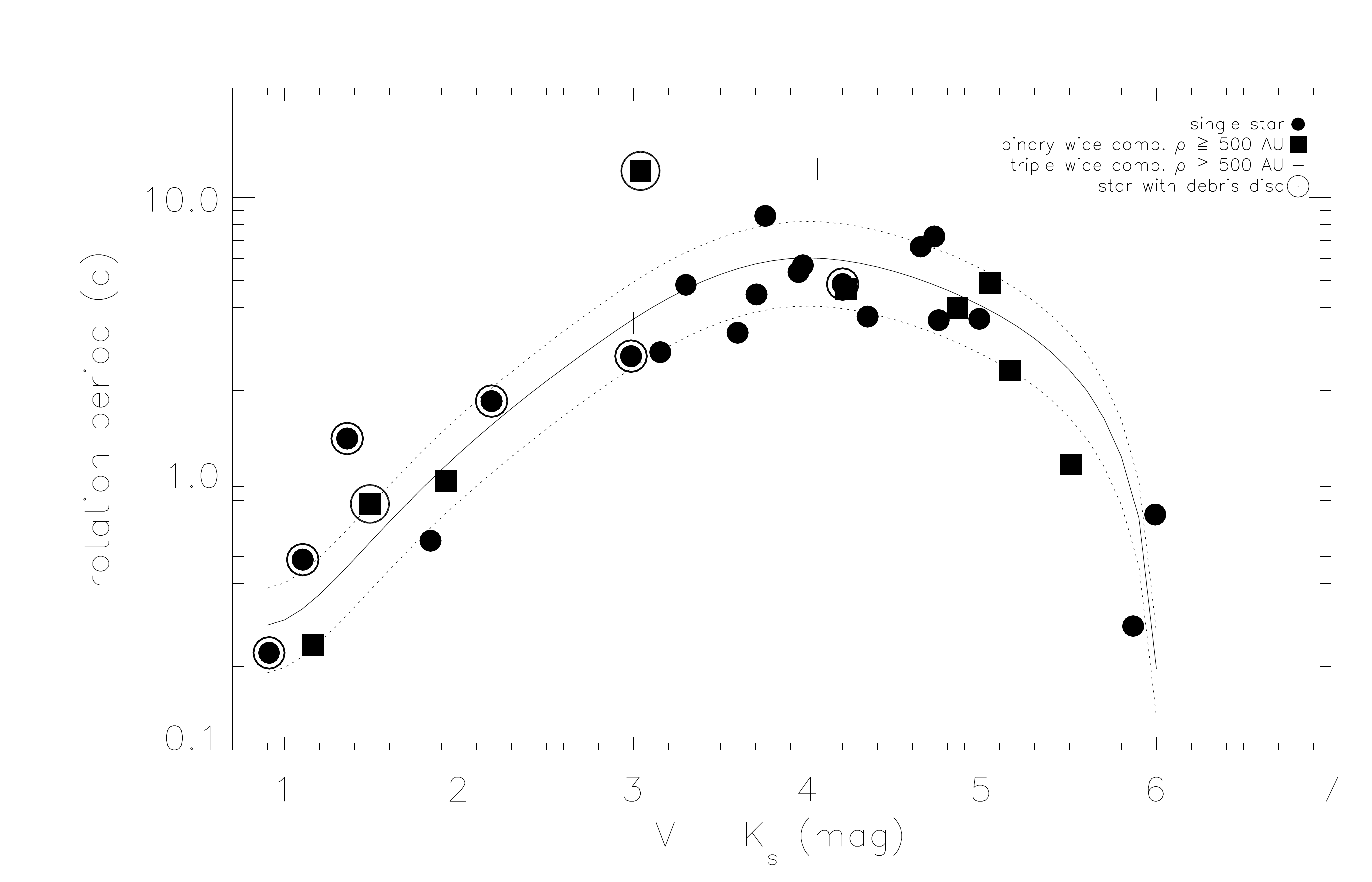}
\end{minipage}
\caption{\label{distri-per_first} Distribution versus V$-$K$_s$ color of the rotation periods of the $\beta$ Pictoris  bona fide members  that are either single (21 stars)  or wide ($\rho$ $>$ 500\,AU) components of binary/multiple systems (14 stars). The meaning of the symbols is given in the legend.  The solid line is a polynomial fit to the rotation periods. Dotted lines represent the $\pm$3$\sigma$ standard deviation of the residuals with respect to the fit.  }
\end{figure*}

\section{Discussion}

It is unanimously accepted that most if not all low-mass stars form with an accretion protostellar disc that, at early stages, magnetically locks the central star 
to an about constant angular velocity (e.g. \citealt{Shu00}). The disc lifetime has a range of values and stars with a long-lived disc reach the Zero Age Main Sequence rotating more slowly than stars with a short-lived disc. Theories predict that the protostellat disc lifetime can be significantly shortened if a binary companion is present, which can truncate the disc, reducing the efficacy of the PMS disc-locking (\citealt{Meibom07}, \citealt{Bouvier93}, \citealt{Edwards93}, \citealt{Ingleby14}, \citealt{Rebull04}), enhancing the mass accretion (\citealt{Papaloizou95}), and finally disrupting the disc (\citealt{Artymowicz92}). In this circumstance, the amplitude of the perturbation should be related to the separation between the components. These predictions are confirmed by observational studies, e.g. by \citet{Kraus16} and \citet{Cieza09}, who found that stars without IR excess tend to have companions at smaller separation than 
stars with excess indicating the presence of a disc. Both studies find that the depletion of protoplanetary discs among binary systems with 
components closer than 40 AU is a factor 2 larger than in either single or wide binaries already at age as young as 1-2 Myr. Moreover, if present, discs around close components ($<$ 30 AU) of binary systems have disc mass depleted by a factor 25 with respect to single stars. The impact of a short-lived discs on rotation in binary systems is also documented by, e.g., \citet{Stauffer16} who report that photometric binaries among the Pleiades GKM-type
stars tend to rotate faster than their counterpart single stars, with an effect that is more pronounced among equal-mass binaries than in single-line
spectroscopic binaries; or by \citet{Douglas16} who report that most, if not all, rapid rotators that deviate from the single-valued relation between mass and  rotation already reached by the age of the Hyades, belong to multiple systems.\\
We are now in the position to extend this investigation of the effect of multiplicity on rotation period distribution towards a much younger age of 25 Myr, using our sample of $\beta$ Pictoris members and candidate members whose single/multiple nature is well characterized. Moreover, in a sparse system like the $\beta$ Pictoris association, one can assume that stellar encounters have a minor role in altering the stellar angular momentum evolution via disc dissipation or enrichment. Moreover, in the absence of nearby massive stars, disc photo-evaporation by external UV radiation can also be ignored. \\
We intend to verify that multiplicity really affects the rotational properties and identify the projected separation at which the components of binary/multiple systems of the $\beta$ Pictoris association start to exhibit rotation periods that significantly deviate from the period distribution of single stars.

 \begin{figure*}
\begin{minipage}{18cm}
\includegraphics[scale = 0.5, trim = 0 0 0 0, clip, angle=90]{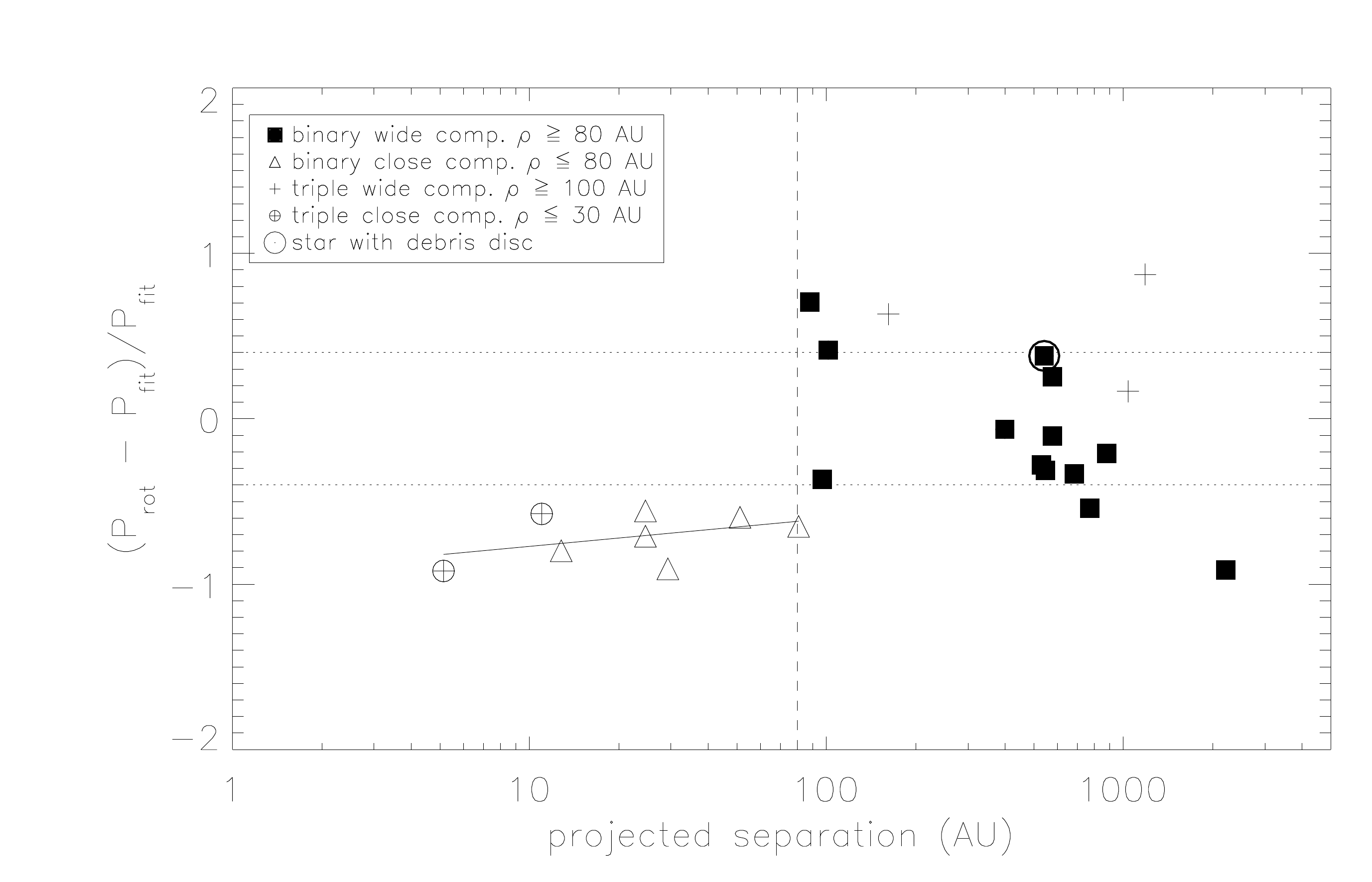}
\end{minipage}
\caption{\label{residuals} Relative residuals versus projected separation (AU) of the rotation periods of all bona fide members in binary/multiple systems with respect to the polynomial fit (solid line in Fig.\,\ref{distri-per_first}). The meaning of the symbols is given in the legend. We note that in our sample there are no components of triple systems with separation in the range 30--100\,AU.}
\end{figure*}
 \subsection{Period distribution of bona fide members}


 \subsubsection{Single stars and components of binary/multiple systems}
We start our analysis considering only  bona fide members that are  single stars and wide components of multiple systems sufficiently distant from each other (projected separation $\rho$ $>$ 500\,AU) to secure that their rotation periods can be considered as they were single stars. In the following, we will show  that separations down to 80\,AU do not affect significantly the observed rotation periods.\\
Then, we selected 35 stars: 21 single stars and 14 wide components of multiple systems with $\rho$ $>$ 500\,AU (Fig.\,\ref{distri-per_first}).
These stars have rotation periods that exhibit the following mass dependence: the rotation period increases towards lower masses (redder colors) reaching a maximum at  V$-$K$_s$ $\simeq$  4\,mag, then decreases towards the very-low-mass regime.  
To measure the mass dependence of the rotation period, we proceeded as follows.  We computed
the median periods over color bins of 1\,mag and computed a polynomial fit to these median values
 (the solid line in Fig.\,\ref{distri-per_first}) valid in the color range 0.9 $<$ V$-$K$_s$ $<$ 6\,mag
	and  whose coefficients are given in Table\,5. 
 We find that the relative residuals with respect to the polynomial fit have a normal distribution with a standard deviation $\sigma$ = 0.11. 	
	The dotted lines represent the $\pm$3$\sigma$ standard deviation from the fit\footnote{We excluded from the fit HIP\,11437A (V$-$K$_s$ =  3.04\,mag; P = 12.5\,d) and HD\,160305 (V$-$K$_s$ = 1.36\,mag; P = 1.341\,d)  because of their
significant  ($>$20$\sigma$)  departure from the general color-period trend.}.
The existence of such a dispersion tells us that the rotation periods of single stars and wide components, in addition to the mass, also depend on other factors such as, for example, differences in the initial rotation periods.\\

 \begin{figure*}
\begin{minipage}{18cm}
\includegraphics[scale = 0.5, trim = 0 0 0 0, clip, angle=90]{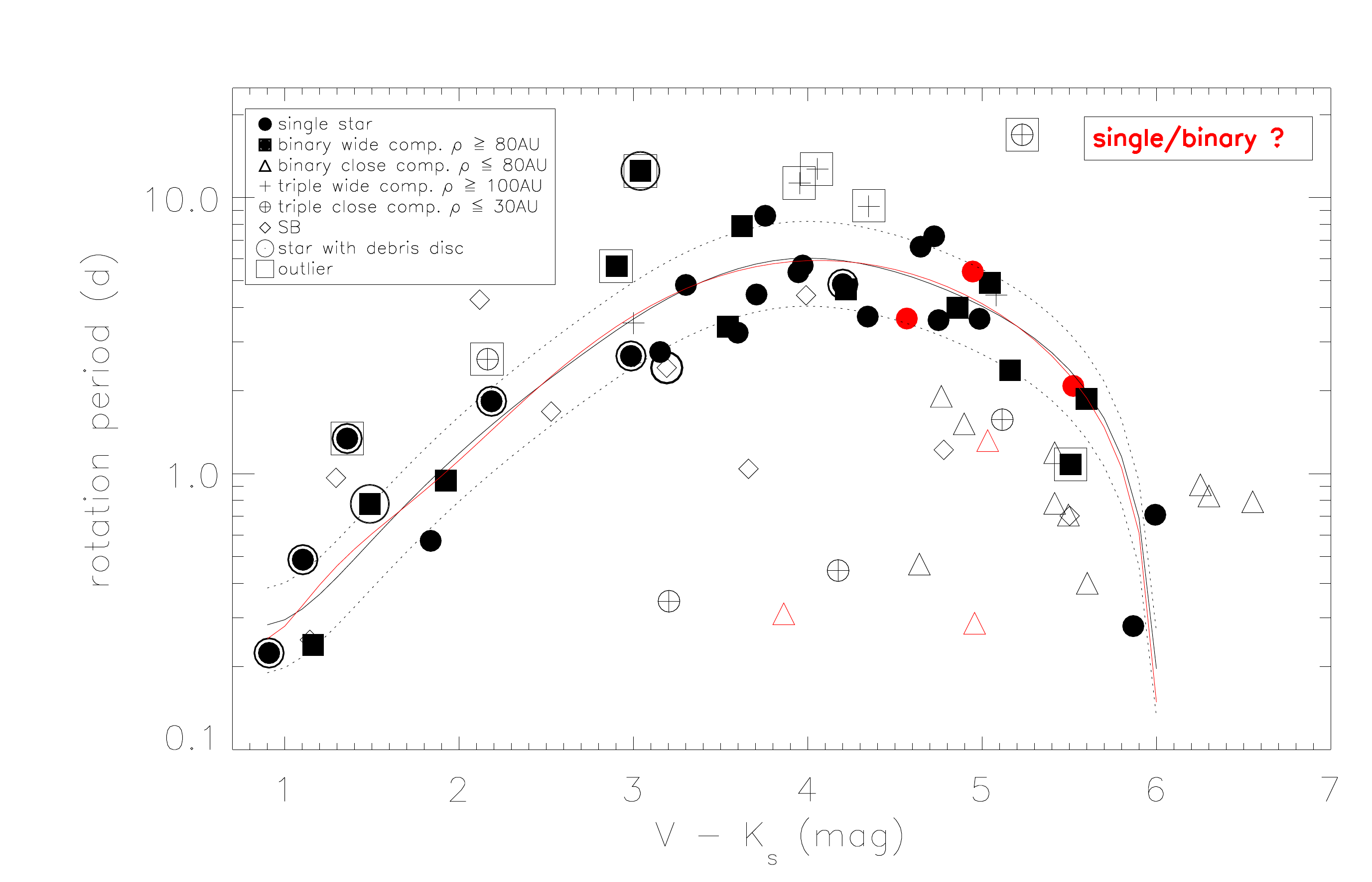}
\end{minipage}
\caption{\label{distri-per} The same as in Fig.\,\ref{distri-per_first}, but with all the bona fide members of the $\beta$ Pictoris association.  Open squares  indicate the members whose rotation periods  significantly deviate either from the general trend exhibited by single stars and components of wide binaries (their residuals from the fit are $>$3$\sigma$) or from the distribution of visual close binaries. Red color is used for members whose single/binary nature is not know. \ We note the segregation of all close binaries at rotation periods shorter than the period distribution of single and wide components of multiple systems. }
\end{figure*}
The fit represents empirically the mass dependence of the rotational period of single stars and wide components of binary/multiple systems. 
The relative residuals (P$_{\rm rot}$ $-$ P$_{\rm fit}$)/P$_{\rm fit}$ with respect to this fit can help us to identify which stars deviate significantly and  to estimate empirically the minimum separation between the components of  a system for which there is no significant departure from this fit. 
These relative residuals  are plotted versus the projected separation (in AU) in Fig.\,\ref{residuals}. 
After excluding single stars and very wide components of multiple systems ($\rho$ $>$ 5000\,AU), and the spectroscopic binaries that will be discussed separately,  we find that the components of multiple systems with a projected separation $\rho$ $\ga$ 80\,AU are mostly within the $\pm$3$\sigma$ distribution of single stars, therefore they behave like they were single stars. On the contrary, all components of multiple systems with a projected separation $\rho$ $\la$ 80\,AU deviate by more than 3$\sigma$. For these residuals, we find a linear Pearson correlation coefficient $r$ = 0.94 with a significance level $\alpha > $ 99.5\%  suggesting that the smaller the separation between the components the faster their rotation period with respect to equal-mass single stars, i.e., their rotation periods are significantly affected/shortened. However, the slope of the linear fit to the distribution of residuals for projected separations $\rho$ $\la$ 80\,AU
 \begin{equation}
 y = -0.87(\pm0.24) + 0.20(\pm0.19)\times log_{10}(\rho)
 \end{equation}
 where $y$ =  (P$_{\rm rot}$ $-$ P$_{\rm fit}$)/P$_{\rm fit}$ (solid line in Fig.\,\ref{residuals}), has a relatively high uncertainty.
 Therefore, owing to the paucity of data so far available, we prefer to be more conservative and to state that all components of multiple systems with a projected separation $\rho$ $\la$ 80\,AU rotate significantly ($>$3$\sigma$) faster, but a linear dependence of rotation rate on separation is only barely detected. \\ 
Among the wide components ($\rho$ $\ga$ 80\,AU) we note four stars\footnote{

TYC\,6878\,0195\,1: V$-$K$_s$ = 2.90\,mag and P = 5.70\,d; BD$-$211074A: V$-$K$_s$ =  4.35\,mag and P = 9.3\,d; TYC\,7443\,1102\,1: V$-$K$_s$ = 3.95\,mag and P = 11.3\,d; TX Psa: V$-$K$_s$ = 5.57\,mag and P = 1.080\,d.
} (all core members) that deviate more than 3$\sigma$ from the general trend exhibited by the majority of stars. Their departure probably indicates that our scenario, where the separation between the components is the dominant parameter that differentiates the period evolution from that of single stars, is a simplification. There are likely other factors that, in individual cases, can be even more important than the separation.\\
In Fig.\,\ref{distri-per}, we plot the rotation periods versus V$-$K$_s$ colors of all bona fide members of the $\beta$ Pictoris association (not only those at $\rho$ $>$ 500\,AU as in Fig.\,\ref{distri-per_first}).  In addition to the  six  mentioned stars  (see footnotes 3 and 4), \rm also 2MASS\,J20013718-3313139 (V$-$K$_s$ = 4.06\,mag; P = 12.7\,d), 2MASS\,J06131330-2742054 (V$-$K$_s$ = 5.23\,mag; P = 16.9\,d), and TYC\,8742\,2065\,1  (V$-$K$_s$ = 2.16\,mag; P = 2.60\,d) deviate significantly from the general color-period trend exhibited by the other members.  The existence of these outliers reminds us that in individual cases other factors apart from mass, component's separation, and initial rotation period, may play a significant role in driving the rotational evolution. \\

 \begin{figure*}
\begin{minipage}{18cm}
\includegraphics[scale = 0.35, trim = 0 0 0 40, clip, angle=90]{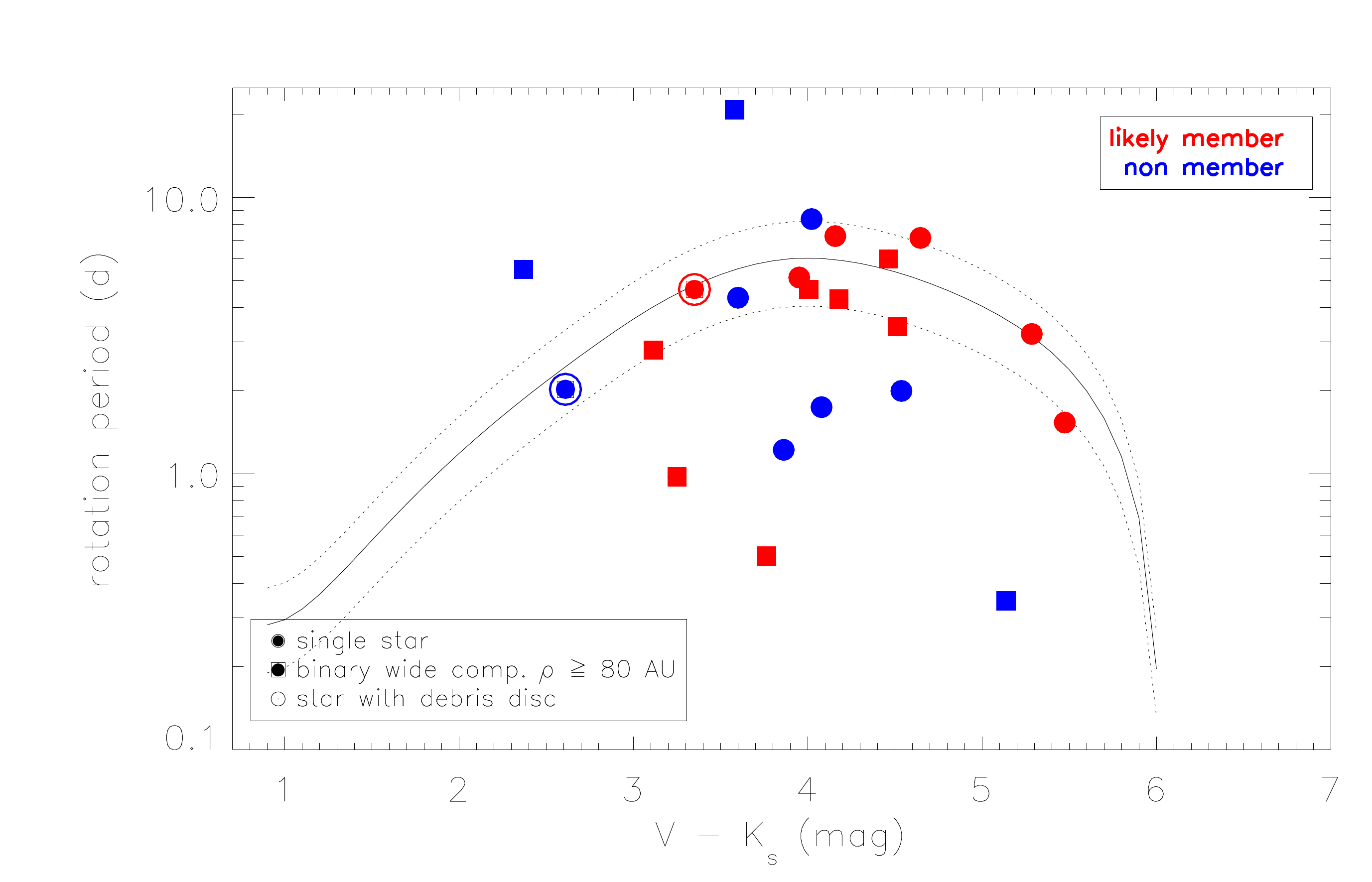}
\includegraphics[scale = 0.35, trim = 0 0 0 40, clip, angle=90]{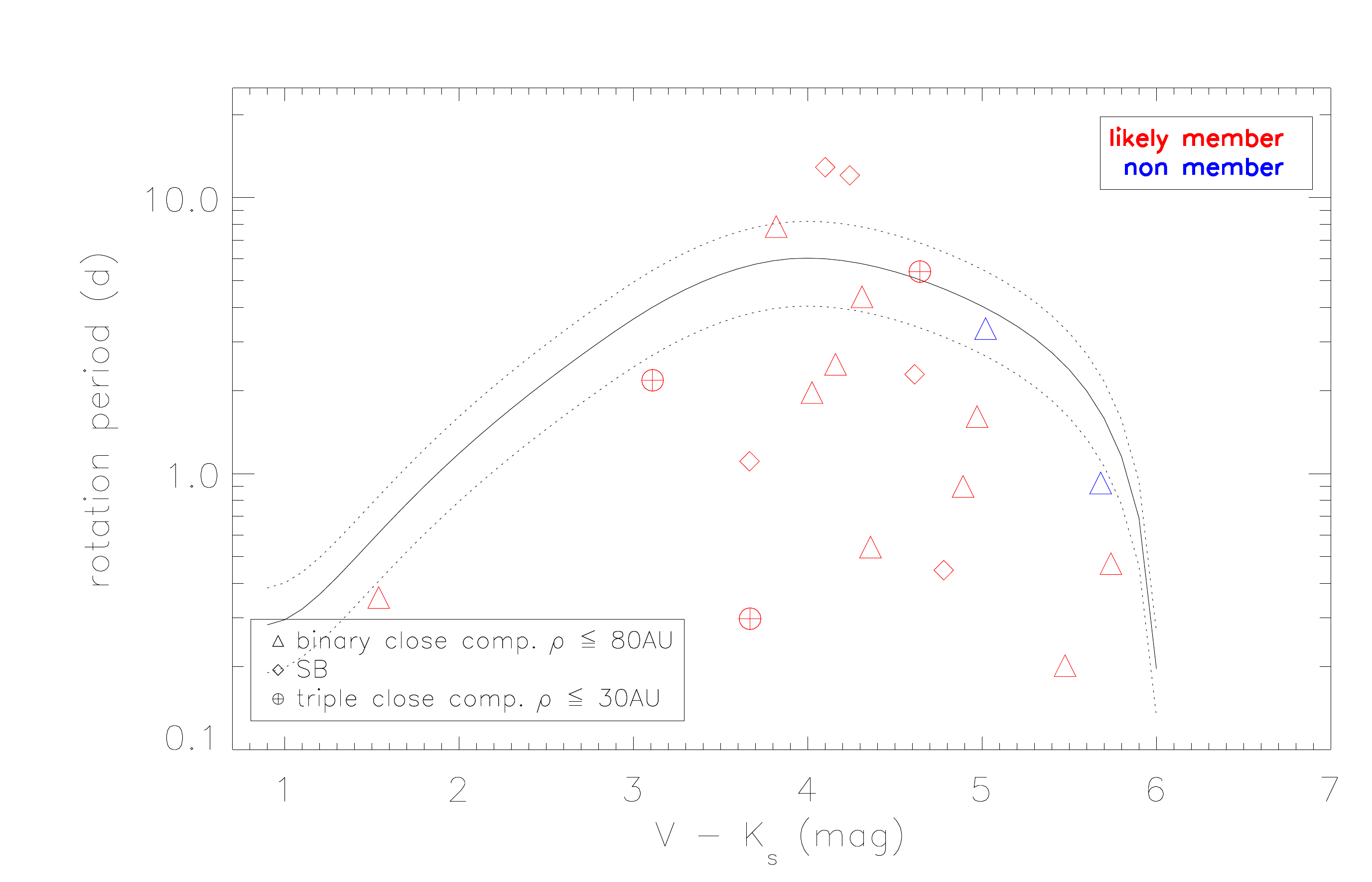}
\end{minipage}
\caption{\label{distri-per_likely} The same as in Fig.\,\ref{distri-per}, but for candidate members (red symbols) of the Association and non members (blue symbols). In the left panel we consider single stars and wide components of binary systems; in the right panel we consider the close components of binary systems and spectroscopic binaries.}
\end{figure*}

\setcounter{table}{4}
\begin{table}
\caption{\label{coeff} Coefficients  and uncertainties  of the polynomial fit to the color-period distribution among bona fide  members that are  single stars and wide  ($\rho$ $>$ 500\,AU) components of multiple systems in the color range 0.9 $<$ V$-$K$_s$ $<$ 6\,mag. }

\begin{tabular}{c@{\hspace{0.1cm}}c@{\hspace{0.1cm}}c@{\hspace{0.1cm}}c@{\hspace{0.1cm}}c@{\hspace{0.1cm}}c@{\hspace{0.1cm}}c@{\hspace{0.1cm}}c@{\hspace{0.1cm}}c@{\hspace{0.1cm}}}
\hline
a0 & a1 & a2 & a3 & a4 & a5 & a6 & a7 & a8 \\
\hline
\tiny 8.38 &\tiny  $-$30.90 &\tiny  46.59 &\tiny  $-$36.37 &\tiny  16.16 &\tiny  $-$4.14 &\tiny  0.60 &\tiny  $-$0.046 &\tiny  0.00140\\
\tiny $\pm$1.75 &\tiny  $\pm$5.09 &\tiny  $\pm$5.99 &\tiny  $\pm$3.73 &\tiny  $\pm$1.36 &\tiny  $\pm$0.30 &\tiny  $\pm$0.04 &\tiny  $\pm$0.002 &\tiny  $\pm$0.00008\\
\hline
\end{tabular}
\end{table}

 \subsubsection{Spectroscopic binaries}

 Our stellar sample totals nine spectroscopic binaries (SBs) that are bona fide members (one of which, TYC\,7408\,0054\,1, is an eclipsing binary). Five SBs have known both the components' separation
and the orbital periods, which are all shorter than 5 days and about synchronized with the rotation period of their primary components (the differences amount to a few percents). The star HIP\,23418 with a rotation period of P = 1.22\,d against an orbital period P = 11.9\,d represents the only exception. Considering the small ($\rho < 0.3$\,AU) component's separation and the about orbital/rotation synchronization, we infer that tidal dissipation has been effective in these stars, as expected from tidal theory (see, e.g. \citealt{Zahn77}; \citealt{Witte02}) and as supported by observational studies (see, e.g. \citealt{Meibom07}). The tidal dissipation makes their angular momentum evolution different from that of single stars or wide components of binary systems. Considering that the remaining four SBs have same age (being bona fide members), similar total masses, and rotation periods shorter than 5 days, we may suppose that they also are likely significantly affected by tidal dissipation. 
Because we are focusing our analysis on effects on angular momentum evolution other than tidal dissipation, and the rotation periods of our SBs are not immediately comparable with those of the other stars in our sample,  all SBs, but HIP\,23418, are excluded from our analysis.\\

We have only three bona fide members in the very-low-mass regime (V$-$K$_s$ $\ge$ 6.0\,mag) that are too red to be compared to the polynomial fit.
This part of our color-period diagram is not enough populated to infer any reliable properties.

Finally, in our sample of bona fide members there are six stars (plotted with red symbols) whose single/binary nature is still
not determined. We note that 2MASS\,J08173943-8243298 (V$-$K$_s$ = 5.03\,mag; P = 1.318\,d), 2MASS\,J17150219-3333398  (V$-$K$_s$ = 3.86\,mag; P = 0.3106\,d), and 2MASS\,J23500639+2659519 (V$-$K$_s$ = 4.96\,mag; P = 0.287\,d) occupy the region of the color-period diagram of close binaries, whereas 2MASS\,05015665+0108429 (V$-$K$_s$ = 5.52\,mag; P =  2.08\,d), 2MASS\,J13545390-7121476 (V$-$K$_s$ = 4.57\,mag; P = 3.65\,d) and 2MASS\,J18420694-5554254 (V$-$K$_s$ =        4.95\,mag; P = 5.403\,d)  occupy the region of the color-period diagram of single stars and wide components.

To summarize, the rotation periods of single stars and wide components of multiple systems with separation $\rho \ga$ 80\,AU exhibit a well defined mass dependence at the age of about 25 Myr that can be approximated by a polynomial fit with a dispersion not larger than a factor two.   Only 9 bona fide members (marked with open squares in Fig.\,\ref{distri-per}) out of 73  (excluding spectroscopic binaries and very-low mass stars)  significantly deviate from the general color-period trend exhibited by the other members.  The rotation periods of close components of multiple systems with separation $\rho \la$ 80\,AU are all shorter and thus populate the region of the color-period diagram below the distribution of single stars and wide components. When the single/binary nature of the cluster or association members is taken into account, the period distribution even at young ages, like the presently considered 25 Myr, has a spread much smaller than claimed in earlier studies.

  \subsection{Period distribution of candidate members}
 The sample of bona fide members has allowed us to discover that single stars and wide components
 of binary/multiple systems have a  period distribution  different than that of components of close binary/multiple systems. 
 We can take advantage of such a different behavior to infer some hint on the candidate members.

  \subsubsection{Single stars}
In our sample there are five single candidate members, which are plotted as red bullets in the left panel of Fig.\,\ref{distri-per_likely}. 
These candidates have some kinematics component larger than 3$\sigma$ but their rotation periods fit well into the color-period distribution of single bona fide members. We consider these stars as likely members of the association. These stars are 2MASS\,J16572029-5343316 (V$-$K$_s$ = 4.65\,mag; P = 7.15\,d), 2MASS\,J23512227+2344207 (V$-$K$_s$ = 5.29\,mag; P = 3.208\,d), 
2MASS\,J16430128-1754274 (V$-$K$_s$ = 3.95\,mag; P = 5.14\,d, which was excluded from the core sample), 
TYC\,5853\,1318\,1 (V$-$K$_s$ = 4.16\,mag; P = 7.26\,d), and 2MASS\,J05294468-3239141 (V$-$K$_s$ = 5.47\,mag; P = 1.532\,d).\\
In the left panel of Fig.\,\ref{distri-per_likely}, we also plot the five single stars that are non members (blue bullets). The rotation periods of three of them deviate significantly from the distribution. However, the rotation periods of  TYC\,915\,1391\,1 (V$-$K$_s$ = 3.60\,mag; P =  4.34\,d)   and 2MASS\,J20055640-3216591  (V$-$K$_s$ =  4.02\,mag; P =  8.368\,d), although non members,   fit well into the distribution. This circumstance poses a severe caveat to the use of the rotation period when inferring the age of individual stars. That is, the fact that the rotation period of a single star fits well into the period distribution of the association is a necessary but not sufficient condition to be classified as member.\\

  \subsubsection{Wide components of binary/multiple systems}
\begin{figure}
\begin{minipage}{10cm}
\includegraphics[scale = 0.35, trim = 0 0 0 40, clip, angle=90]{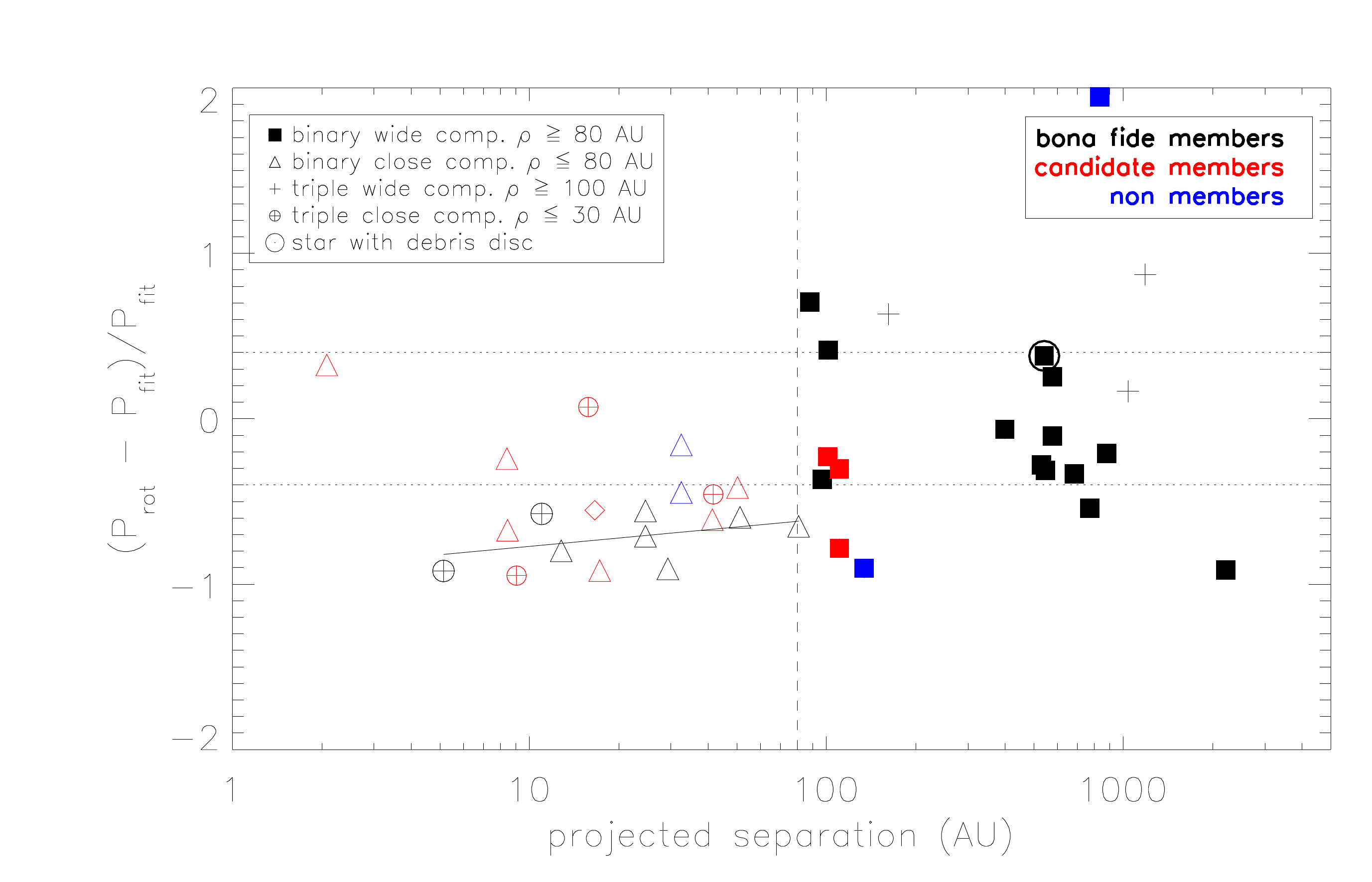}
\end{minipage}
\caption{\label{residuals_tot} The same as in Fig.\,\ref{residuals}, but with inclusion of candidate members and non members.}
\end{figure}

Similarly, in our sample there are  seven candidate members that are wide components of multiple systems (red squares). 
These candidates have some kinematics component larger than 3$\sigma$ but the rotation periods of six of them fit well into the color-period distribution of single bona fide members. We consider these stars as likely members of the association.  These stars are 2MASS\,J02014677+0117161   (V$-$K$_s$ = 4.51\,mag; P = 3.41\,d), and RBS\,269 (V$-$K$_s$ = 4.46\,mag; P =  6.0\,d ), which were excluded from the core sample,
2MASS\,J04435686+3723033   (V$-$K$_s$ = 4.18\,mag; P = 4.288\,d),
 2MASS\,J18202275-1011131A   (V$-$K$_s$ = 3.35\,mag; P = 4.655\,d),
 2MASS\,J18202275-1011131B   (V$-$K$_s$ = 4.01\,mag; P = 5.15\,d), and
TYC\,1208\,0468\,1    (V$-$K$_s$ = 3.11\,mag; P = 2.803\,d). For this star, however, we note that 
the rotation period is shorter than that of other members with similar separation ($\sim$100\,AU) (see also Fig.\,\ref{residuals_tot}).\\
The exception is represented by BD+262161B   (V$-$K$_s$ = 3.25\,mag; P = 0.974\,d) and TYC\,6872\,1011\,1   (V$-$K$_s$ = 3.76\,mag; P =  0.503\,d, which was excluded from the core sample)
whose rotation periods are in disagreement with the distribution.

In the left panel of Fig.6, we also plot the five non member wide components (blue squares): the following stars have accordingly their rotation period in disagreement with the distribution, that is 2MASS\,J01365516-0647379   (V$-$K$_s$ = 5.14\,mag; P =  0.346\,d);  HIP105441   (V$-$K$_s$ = 2.37\,mag; P =    5.50\,d); and  TYC\,9114\,1267\,1   (V$-$K$_s$ = 3.58\,mag; P = 20.8\,d). An exception is represented by the debris disc BD\,+26\,2161A   (V$-$K$_s$ = 2.61\,mag; P = 2.022\,d) whose  rotation period is in agreement with the distribution.

  \subsubsection{Close components of binary/multiple systems}
In the right panel of Fig.\,\ref{distri-per_likely}, we plot the 9 components of close binary systems that are candidate members (red triangles). They all but  one exhibit rotation periods that are below the distribution of single and wide components, similarly to close components  that are  bona fide members. However, differently than single stars, this information is not a strong constraint to the membership. For these stars we can state that their rotation period is in agreements with the distribution of the members. However, this is a necessary but not sufficient condition to be considered members. There are other association/clusters with different ages whose period distribution significantly overlap at these short rotation regimes. The only exceptions are HIP\,50156 and BD\,-21\,1074B whose rotation periods are too long with respect to the close components members of the $\beta$ Pictoris association.  We plot also the three  close binary non members (blue triangles), which exhibit rotation periods that tend to be too long and in disagreements with the distribution of the close binary members.
    
\subsubsection{Spectroscopic binaries}
 Our stellar sample totals four spectroscopic binaries  that are candidate members  (red diamonds). We have information neither on orbital period nor on component's separation. Also their age is not definite, according to their candidate status. Therefore, we are not in position to infer if
their angular momentum has suffered or not significant tidal dissipation. As in Sect.\,5.1, we exclude them from our analysis. \\

In Fig.\,\ref{residuals_tot}, we plot the period residuals with respect to the fit versus the projected separation, as in Fig.\,\ref{residuals}, of all bona fide members, candidate members and non members. As shown by the red color, those stars that we classified as likely members follow the distribution exhibited by bona fide members.
Our investigation, therefore, supports their candidate membership of the $\beta$ Pictoris association.
On the contrary, the following candidate members:  HIP\,50156 ($\rho$ = 2.08\,AU; (P$_{\rm rot}$ $-$ P$_{\rm fit}$)/P$_{\rm fit}$ = 0.325), BD\,-21\,1074B ($\rho$ = 15.79\,AU; (P$_{\rm rot}$ $-$ P$_{\rm fit}$)/P$_{\rm fit}$ = 0.071), BD\,+26\,2161B  ($\rho$ = 110.7; (P$_{\rm rot}$ $-$ P$_{\rm fit}$)/P$_{\rm fit}$ = -0.78); TYC\,4770\,0797\,1  ($\rho$ = 8.4\,AU; (P$_{\rm rot}$ $-$ P$_{\rm fit}$)/P$_{\rm fit}$ = -0.24),  do not follow the general trend and  our study suggests their non membership.

 To summarize,  the rotation period represents a valuable information when assessing the membership of a star provided that its single/multiple nature and, in the latter case, the separation between the components, are known.  The good fitting of the rotation period into the distribution of a proposed association/cluster is a necessary condition for the star to be member, although the rotation period alone does not provide a sufficient condition. 
 
    \subsection{Wide components of triple systems}
We note that the wide components of triple systems tend to have rotation periods comparable to but slower than either single
stars or components of wide binaries. It seems that in these multiple systems, the initial angular momentum of the 
protostellar cloud, since divided among more components, may have given a fraction of it to the wide component, at least,
smaller than what happens in case of binary systems.  \\

The analysis presented in this Section shows that, when studying the stellar angular momentum evolution using the rotation period distributions of associations and open clusters, it is fundamental to know the single/binary nature of each member since their
rotational properties are significantly different. Mixing single/wide components of multiple systems together with components of close binaries has the consequence to mask the real period segregation between these two different classes of stars, to make the rotation spread to appear larger than it is, and to bias the mean/median/percentile periods of a given cluster towards smaller values.
\begin{figure*}
\begin{minipage}{20cm}
\includegraphics[scale = 0.3, trim = 0 0 0 0, clip, angle=90]{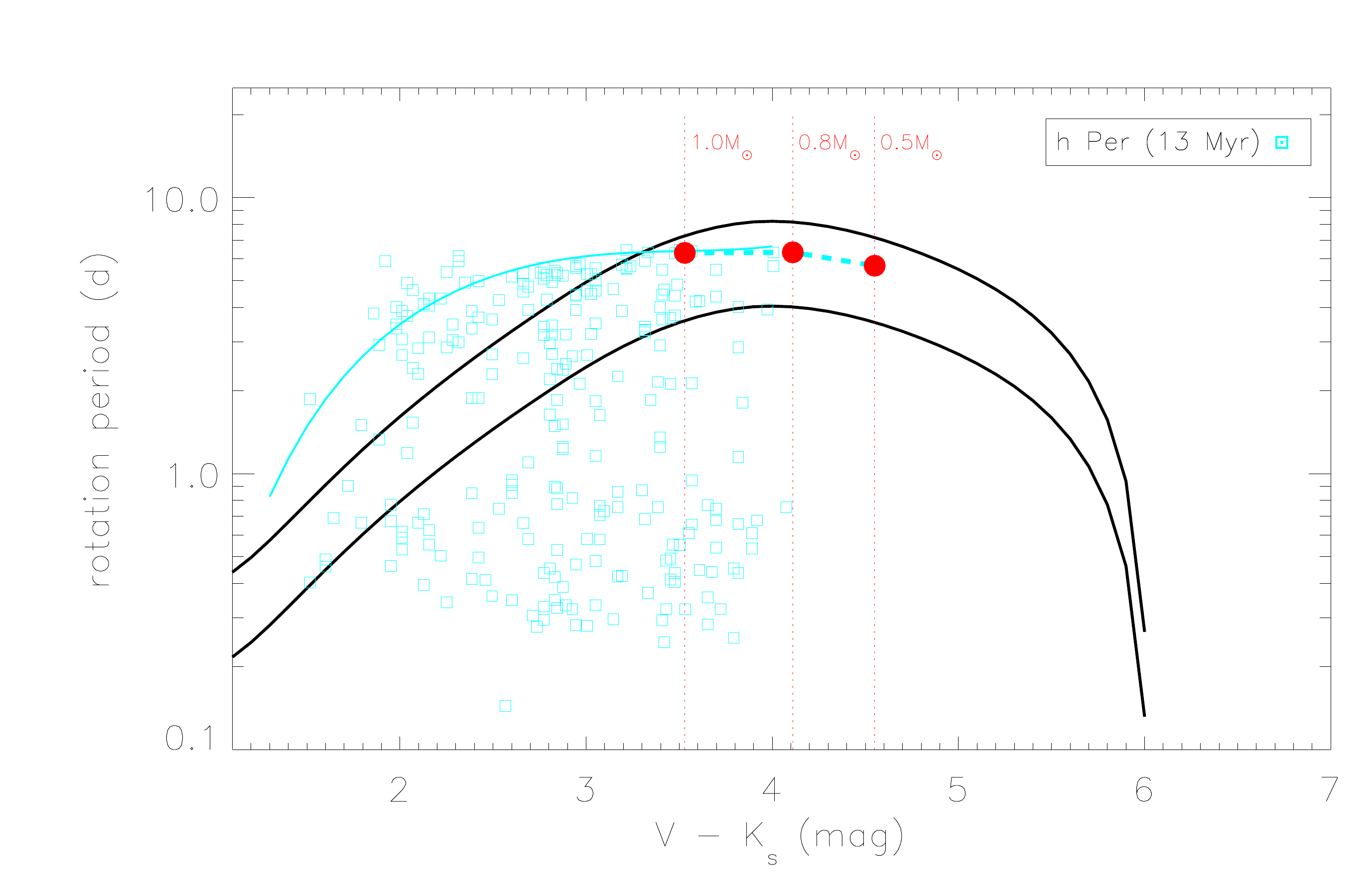}
\includegraphics[scale = 0.3, trim = 0 0 0 0, clip, angle=90]{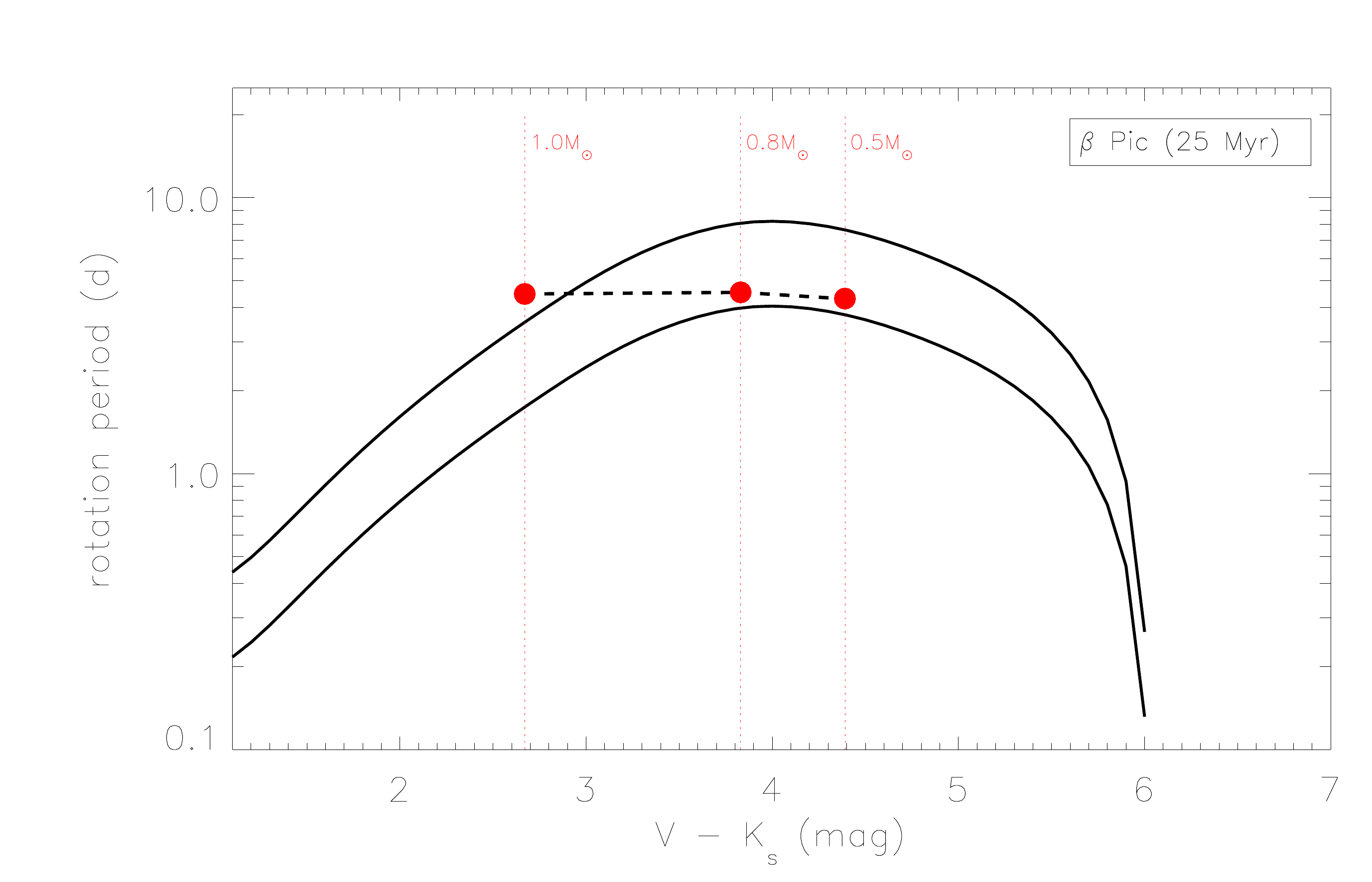}\\
\includegraphics[scale = 0.3, trim = 0 0 0 0, clip, angle=90]{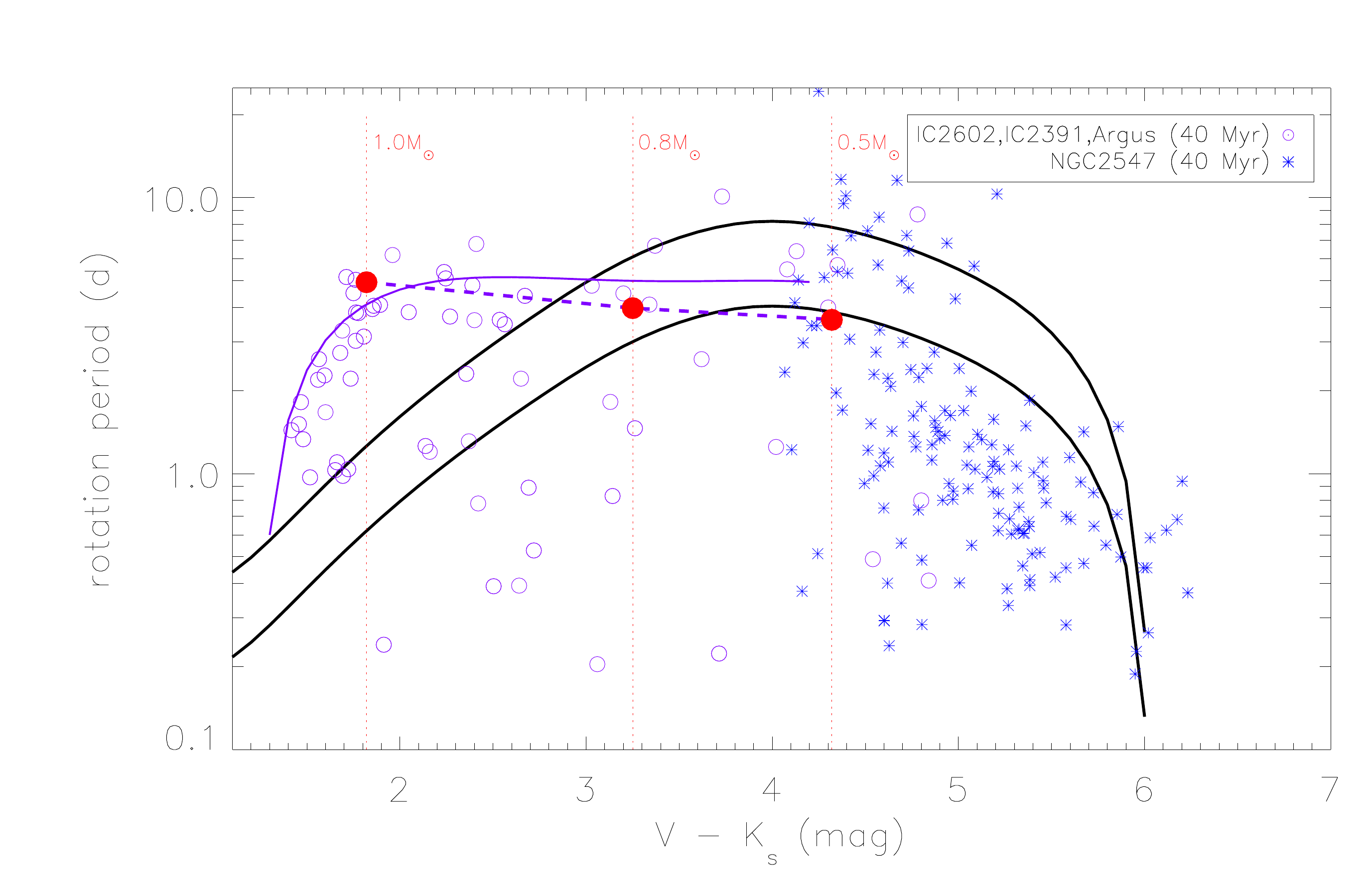}
\includegraphics[scale = 0.3, trim = 0 0 0 0, clip, angle=90]{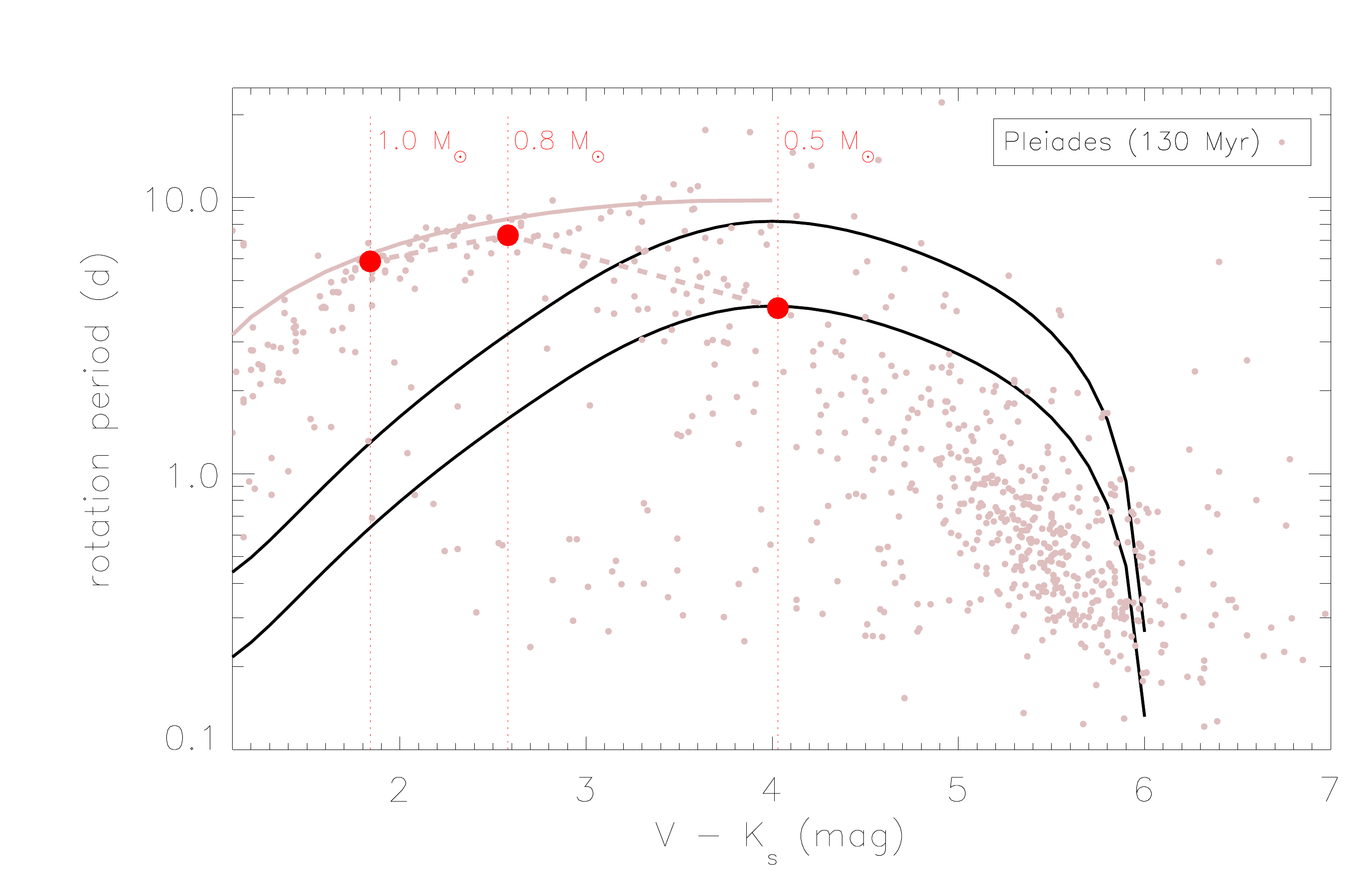}
\end{minipage}
\caption{\label{comparison} The rotation period distribution of the $\beta$ Pic members (only the $\pm3\sigma$ fits to the distribution are plotted as solid black lines) is compared with the distribution of the $h$ Per single members (light-blue open squares) in the top panel; with the distribution of  IC\,2391+IC\,2602+Argus members (violet open circles),  and NGC\,2547 members (blue asterisks) in the middle panel, and with the distribution of the Pleiades members (brown small bullets) in the bottom panel. Thick solid lines are linear fits to the 90th percentile of the  distribution of the comparison clusters. Filled bullets connected with dashed lines indicate the model rotation period predicted by the \citet{gallet15} for 1.0, 0.8 and 0.5 M$_{\odot}$.}
\end{figure*}

\section{Comparison with other open clusters/associations}
A study of the rotation period distribution of the $\beta$ Pictoris members in the context of the angular momentum evolution is out of the
scope of this paper. Nonetheless, a comparison  with the rotation period distribution of other associations/clusters of different ages
is very useful to infer some preliminary and qualitative results, at least, to be further developed elsewhere. \\
The more recent studies (see, e.g., \citealt{Messina16a}) point towards ages of the $\beta$ Pictoris association from 21 to 26 Myr. These estimates are significantly older than the estimates made by, e.g., \citet{Zuckerman01} and \citet{Song03}, but, for instance, closer to the very first estimate provided by \citet{Barrado99}.\\
The open cluster $h$ Persei (\citealt{Moraux13}) with an age of about 13 Myr 
and the open clusters/associations NGC\,2547 (\citealt{Irwin08}), IC\,2391, Argus (\citealt{Messina11}), and IC\,2602 (\cite{Barnes99}) with an age of about 40 Myr have the closest ages to that of $\beta$ Pictoris 
and have known rotation period distributions. Unfortunately, we face three major limits when comparing their rotation period distributions. First, the single/binary nature of the comparison cluster and association members is not known as accurately as for the $\beta$ Pictoris members. For this reason, we will limit the comparison to the upper envelopes of the period distributions, which are likely represented by single stars and wide-orbit binaries. Second, we know the rotation periods of only  the 
higher mass members (1 $<$ (V$-$K$_s$)$_0$ $<$ 4\,mag) of $h$ Persei.  Therefore, at the lower mass regime the comparison is possible only at three time steps (25, 40, and 130\,Myr.)  Finally, loose and very sparse associations, as $\beta$ Pictoris, and open clusters  may  represent
two different environments for the dynamical evolution of their members,  in the sense that  effects of binary encounters on the primordial disc lifetime and, therefore, on the early rotational evolution may be different in magnitude (see, \citealt{Clarke93}, \citealt{Heller95}). \rm Therefore, any results should take into account this major difference.\\
To compare the distributions, we correct the V$-$K$_s$ color of the $h$ Per members for interstellar reddening comparing masses and colors taken from Moraux et al. (2013)  with the  V$-$K$_s$ versus mass relation for young stars from \cite{Pecaut13}. As a check, we find that the average color excess E(V$-$K$_s$) = 1.60\,mag derived with our approach  is in good agreement with the E(V$-$K$_s$) = 1.52\,mag inferred from E(B$-$V) = 0.54\,mag (\citealt{Mayne08}), assuming R$_{\rm V}$ = 3.1. Although we have indication on which members of $h$ Persei are photometric binaries, we do not know the projected separation of their components, therefore we have no possibility to distinguish close from wide orbit binaries, as we did for the $\beta$ Pictoris members.  For this reason, we focus on only the single members of $h$ Persei. \\ \rm 
We proceed similarly with the NGC\,2547 members using the colors and masses provided by \citet{Irwin08}. However, in this case we infer an average color excess E(V$-$K$_s$) = 0.53\,mag significantly larger than E(V$-$K$_s$) = 0.20\,mag
 derived from the E(B$-$V) = 0.06\,mag (\citealt{Irwin08}). The reason of this discrepancy is not clear to us. However, irrespectively from the use of the smaller or larger reddening correction, when we overplot the rotation period distribution of the NGC\,2547 members on the rotation period distribution of the $\beta$ Pictoris bona fide members, we find qualitatively the same result. For the NGC\,2547 members we have no indication on their single or binary nature.\\
 The V$-$K$_s$ colors of the  IC\,2391, IC\,2602, and Argus members were derived using 
 the  K$_s$ magnitudes from 2MASS catalog (\citealt{Cutri03}) and V magnitudes from \cite{Messina11} and \cite{Barnes99}.\\
  The results of the comparison are summarized in the panels of Fig.\,\ref{comparison}. 
The comparison period distributions clearly exhibits fast and slow rotators. We use the 90th percentiles computed in 0.5-mag color bins to identify the upper envelope of the rotation period distributions of the comparison cluster and associations. These are represented with heavy solid lines (light-blue for $h$ Persei in the top panel, violet for IC\,2391+IC\,2602+Argus in the middle panel, and brown for Pleiades in the bottom panel) that mark the position of the slowest members.\\ 
The slow F-G members of $h$ Persei and of IC\,2391+IC\,2602+Argus rotate significantly slower than the F-G members of $\beta$ Pictoris.  Using the known scenario of PMS angular momentum evolution as guideline, we can infer from Fig.\,\ref{comparison} that F-G stars at an age of about 13 Myr ($h$ Persei members) are still spinning up, owing to radius contraction and angular momentum conservation. They reach a likely maximum rotation rate at an age of about 25 Myr ($\beta$ Pictoris members), then after start to slow down, owing to the combined effect of rotation braking by magnetized stellar winds and core-envelope decoupling (see, e.g. \citealt{Spada11}),  gaining by the age of about 40 Myr the position in the color-period diagram occupied by the IC\,2391+IC\,2602+Argus slow members. The rotation magnetic braking keeps going with age as shown, for comparison,  by the Pleiades members (bottom panel; \citealt{Rebull16}) at an age of about 130 Myr. Such an observational pattern is predicted quite well by models of angular momentum evolution for 1.0\,M$_{\odot}$ and 0.8\,M$_{\odot}$ stars. In  Fig.\,\ref{comparison} we plot the \citet{gallet15} model rotation periods at the sampled ages    with filled bullets connected by dashed lines. Vertical dotted lines indicate the V$-$K$_s$ colors corresponding to 1.0\,M$_{\odot}$, 0.8\,M$_{\odot}$, and 0.5\,M$_{\odot}$ derived from the \citet{Baraffe08} models used by \citet{gallet15}. Some level of disagreement exists for  0.8\,M$_{\odot}$ stars at 25\,Myr and 40 Myr, where model periods are shorter than observed.  \\ \rm
Among the slow mid-K to early-M stars it is more complicated to retrieve the angular momentum evolution pattern since these stars have distributions that apparently do not differ significantly from each other at the three time steps  25 Myr, 40 Myr and 130 Myr, giving some hint that the angular momentum evolution of mid-K to early-M stars has been negligible in the 25--130 Myr time interval. The models of angular momentum evolution actually predict a monotonic increase of the rotation rate only from  13 to 40 Myr, and about a constant rotation period up to 130 Myr for the 0.5\,M$_{\odot}$ stars (see, e.g. \citealt{gallet15}). \\
 Finally, among the mid- to late-M stars we note that the rotation period distributions at the 40 Myr and 130 Myr steps are about indistinguishable and their upper envelope, consisting of the slow rotators, is below the distribution of the $\beta$ Pictoris members.  We can interpret this result assuming that mid- to late-M stars undergo the stellar radius contraction until about the age of the Pleiades and, therefore, they are observed to spinning up their rotation period from the age of $\beta$ Pictoris until the Pleiades age, when models of angular momentum evolution predict these stars to reach the maximum rotation rate.  \\
 The result that we found for the F and G stars  is very important and should be kept in mind when using  the rotation period as age indicator. In fact,
 we found that in the age range from $\sim$13\,Myr to $\sim$40\,Myr the dependence of the rotation period of either single stars or wide components of multiple systems on age is ambiguous.  Stars with ages in the 13--25\,Myr range (when periods are spinning up) are expected to have again a similar period in the  25--40\,Myr range (when periods are slowing down). The uncertainty on the age determination of F--G stars will be minimum at 25\,Myr and progressively larger as far as we move to younger or older  (up to 40\,Myr) ages. On the other hand, such a kind of degeneracy in the age estimate can be successfully removed when the complementary information on the Li EW is available. 
 
\section{Light curve amplitude versus rotation}

The photometric rotational modulation exhibited by all targets arises from the presence of spots unevenly distributed
along their stellar longitudes. The peak-to-peak amplitude of this modulation depends on
the spot's area, its temperature contrast with respect to the unspotted photosphere, the photometric band, and on a combination of  average latitude where  spots are located and the inclination of the stellar rotation axis 
with respect to the observer's line of sight. These last quantities can play in reducing the observed amplitude for a fixed spot area and temperature contrast.\\
Moreover, the light curve amplitude generally changes versus time on the same star  due to active region growth and decay, latitude migration, and presence of spot cycles and/or long-term trends.\\
This is the reason why stars of similar masses, rotation periods, and ages show a distribution of amplitudes.
The amplitude can be then used as an indicator of a lower limit to the level of activity hosted by the star and, when a series of amplitude measurements are available for a given star, the largest value better represents the maximum activity level that the star can exhibit (see, e.g., \citealt{Messina01}, \citeyear{Messina03}).\\

\begin{figure}
\begin{minipage}{10cm}
\includegraphics[scale = 0.55, trim = 0 0 0 0, clip, angle=0]{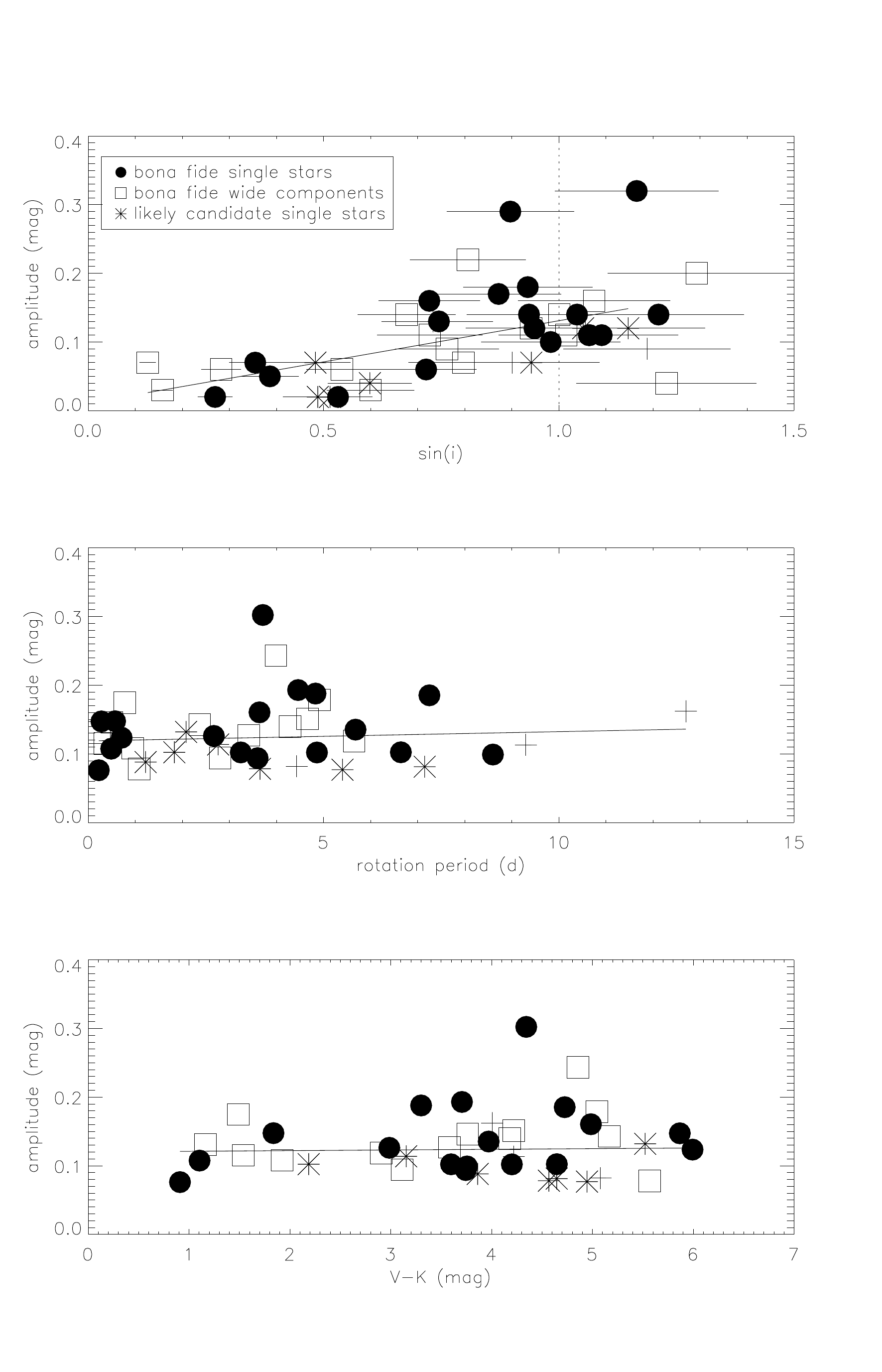}
\end{minipage}
\caption{\label{distri_sini} Top panel: Distribution of V-band light curve amplitudes versus $\sin{i}$ for bona fide  members that are  single stars (bullets), wide components of multiple systems (open squares), and single likely candidate members.  Middle panel: same as in the top panel but with amplitudes decorrelated from $\sin{i}$. Bottom panel: Distribution of decorrelated amplitude versus V$-$K$_s$ color. Solid lines in all panels  represent linear fits.}
\end{figure}

 The correlation between light curve amplitude and rotation period was earlier investigated in $\beta$ Pictoris members by Messina et al. (\citeyear{Messina10}, \citeyear{Messina11}) who  found no significant correlation. However, the number of available amplitude measurements for each association was not large as in the present case of the $\beta$ Pictoris association. Moreover, in those studies no distinction was made between single/wide components of multiple systems and components of close binaries.\\
In Fig.\,\ref{distri_sini}, the light curve amplitudes of the $\beta$ Pictoris members are plotted versus  $\sin{i}$.
Light curve amplitudes, stellar radii, rotation periods, and projected rotational velocities used to derive $\sin{i}$ are all taken from Paper II. We find that the candidate members (plotted with asterisks) that are single or components of wide binaries, and that were found in our previous analysis to have rotation periods that well fit into the period distribution of bona fide members, have a distribution of amplitudes indistinguishable from that of bona fide members. These stars will be also considered in the following analysis. Light curve amplitudes are measured from the amplitude of the sinusoidal fit to the phased light curves. We find with a Kolmogorov test that single stars and wide components of binary/triple systems exhibit the same distribution. This circumstance further confirms that wide components of multiple systems behave as single stars also on the photometric variability point of view. \\
From the top panel of Fig.\,\ref{distri_sini}, we infer that the amplitude is positively correlated to the $\sin{i}$ with a Spearman rank correlation $\rho$ =   0.53 and $p$-value 10$^{-3}$. This result is expected since equator-on stars ($\sin{i}$ = 1) maximize the amplitude of the rotational modulation of starspots with respect to low-inclination ($\sin{i}$ $<$ 1) stars. \\
We can use the linear fit (solid line) to remove the effect of inclination on the amplitude distribution and compute new amplitudes as all stars were equator-on. Then, in the middle panel of Fig.\,\ref{distri_sini}, we plot these inclination-corrected amplitudes versus rotation period. We find a Spearman rank correlation $\rho$ =   0.06 and $p$-value $\sim$ 0.50 that allows us to conclude that the amplitude is not correlated to the rotation. This is a very different behavior with respect to older stars, like the AB Doradus and the Pleiades members (see, \citealt{Messina01}, \citeyear{Messina03}) whose light curve amplitudes are strongly and negatively correlated to the rotation period. We still note a significant dispersion of the amplitudes around their mean value.\\
In the bottom panel of Fig.\,\ref{distri_sini}, we investigate the dependence of the light curve amplitude on the color, i.e. on the stellar mass. Again we find no correlation with a Spearman rank correlation $\rho$ =   0.02 and $p$-value $\sim$0.30. Again, the amplitudes show a level of dispersion that we attribute to the variable level of activity with time. We note an increases of dispersion, with the highest values around K and early-M stars.\\
 Similar results are reported by \cite{Moraux13} for the $h$ Per cluster at the age of 13 Myr. They find the light curve amplitudes to be uncorrelated to the rotation period. Rather, a week dependence on mass is found, with the lower mass stars to have light curve amplitudes slightly larger than higher mass stars. Similarly at the older age of 40 Myr, the light curve amplitudes of the NGC\,2547 members still appear to be uncorrelated to the rotation period (\citealt{Irwin08}). 
Three single stars TYC\,915\,1391\,1 (with no $v\sin{i}$),  TYC\,9073\,0762\,1, and 2MASS\,J21100535-1919573 all have amplitudes  significantly larger than the average ($>$0.29\,mag). These certainly deserve additional study.

\section{Conclusions}

We have assessed the membership of  the $\beta$ Pictoris association members using Galactic velocity (UVW) and space (XYZ) components derived from updated values of proper motions, radial velocities, and distances, complemented with information on Li content, and rotation period. As result, we have identified 80 bona fide members, 22 candidate members, and 15 non members on a total of 117 stars.\\

Analyzing the sample of bona fide  members, we found that single stars and components of multiple systems with separation larger than about 80\,AU  have the same distribution of rotation periods vs. the V$-$K$_s$ color. 
On the contrary, components of close visual binaries/triples with separation smaller than about 80\,AU  rotate preferentially faster than their equal-mass single counterparts. This circumstance suggests that when the components are sufficiently close,  their primordial discs undergo an enhanced dispersal allowing the stars to start their spin up earlier that single stars. \\
The characterization of the period distribution made by us and based on bona fide members, has allowed us to infer additional information on candidate members whose single/binary nature is known. As result of this comparison, we find that among our candidate members 17 stars (five single stars, six wide orbit components, and eight close orbit components of multiple system) have rotation periods that further support their membership. On the contrary, three candidate members (one single star  and two close components of multiple systems) have rotation periods that favor their non membership.\\
All but one spectroscopic binaries in our sample have rotation periods that are not immediately comparable with those of either single/wide components or close components of multiple systems since they likely suffered significantly from tidal effects.\\
A comparison with the rotation period distributions of the younger $h$ Persei open cluster ($\sim$13\,Myr)  and the older 40-Myr IC\,2391, IC\,2602, Argus, NGC\,2547 and the 130-Myr Pleiades members shows that F and G stars at the age of 13 Myr have not reached yet the zero-age-main-sequence and, therefore, are still contracting their radius spinning up their rotation. They reach a likely maximum rotation rate at the age of about 25 Myr (represented by the $\beta$ Pictoris members).  Subsequently, they start a monotonic rotation slowing down which, in our comparison,  is readily visible until the Pleiades age. 
This is the scenario also predicted by models of angular momentum evolution. Differently than model prediction, the K and early-M stars in our sample exhibit period distributions that are apparently indistinguishable from each other. That means that in this mass range the  single and wide components $\beta$ Pictoris members apparently have rotation periods similar to those of either younger or older stars. \rm However, this mass range in our comparison is not represented as significantly as the F and G mass range. Finally, mid- to late-M stars older than 25 Myr all appear to rotate significantly faster than the $\beta$ Pictoris members,  giving hint that the rotation spinning up is proceeding in this mass range. 
Finally, we find that the distribution of  light curve amplitudes of single stars is undistinguishable from that of wide components of multiple systems. Moreover, the amplitude is found to increase with $\sin{i}$, as expected from geometrical considerations. After decorrelating the dependence on $\sin{i}$, we found no dependence of the amplitude on the rotation period.\\

{\it Acknowledgements}. Research on stellar activity at INAF- Catania Astrophysical Observatory 
is supported by MIUR  (Ministero dell'Istruzione, dell'Universit\`a e della Ricerca).  This research has made use of the Simbad database, operated at CDS (Strasbourg, France).   
LZ acknowledges support by the Joint Research Fund in Astronomy (U1431114 and U1631236) under cooperative agreement between the National Natural Science Foundation of China and Chinese Academy of Sciences.  We are very grateful to the Referee whose valuable comments helped us to significantly improve the quality of the paper.
\\

\bibliographystyle{aa.bst} 
\bibliography{mybib} 

\clearpage
\setcounter{table}{0}

\begin{deluxetable}{c@{\hspace{0.05cm}}rr@{\hspace{0.05cm}}r@{\hspace{0.05cm}}r@{\hspace{0.05cm}}c@{\hspace{0.05cm}}rrrr}

\tablecolumns{10}
\tablecaption{\label{tab_period} List of $\beta$ Pictoris members analysed in this study: name, RA and DEC coordinates, V mag, V$-$K$_s$ color, spectral type, rotation period and its uncertainty, light curve amplitude, and info on binarity.
}
\tablewidth{0pt}
 \tablehead{\colhead{Target} & \colhead{RA } & \colhead{DEC} & \colhead{V} &  \colhead{V$-$K} & \colhead{Sp.T} & \colhead{P} & \colhead{$\Delta$P} & \colhead{$\Delta$V}  & \colhead{type} \\
\colhead{    } & \colhead{(J2000)} & \colhead{(J2000)} & \colhead{(mag)} & \colhead{(mag)} & \colhead{} & \colhead{(d)} & \colhead{(d)} & \colhead{(mag)} & \colhead{  }\\
}
\startdata

  \hline
\tiny HIP\,560  &  00 06 50.08  &  -23 06 27.20  &  6.15  &  0.91  &  F3V  &  0.224  &  0.005  &  0.008  &  S+D\\
\tiny 2MASS J00172353-6645124  &  00 17 23.54  &  -66 45 12.50  &  12.35  &  4.65  &  M2.5V  &  6.644  &  0.027  &  0.100  &  S\\
\tiny TYC\,1186\,0706\,1  &  00 23 34.66  &  20 14 28.75  &  10.96  &  3.62  &  K7.5V+M5  &  7.9  &  0.1  &  0.070  &  Bw\\
\tiny GJ\,2006A  &  00 27 50.23  &  -32 33 06.42  &  12.87  &  4.86  &  M3.5Ve  &  3.99  &  0.05  &  0.170  &  Bw\\
\tiny GJ\,2006B  &  00 27 50.35  &  -32 33 23.86  &  13.16  &  5.04  &  M3.5Ve  &  4.91  &  0.05  &  0.120  &  Bw\\
\tiny 2MASS J00323480+0729271A  &  00 32 34.81  &  07 29 27.10  &  13.40  &  5.02  &  M4V  &  3.355  &  0.005  &  0.045  &  Bc\\
\tiny 2MASS J00323480+0729271B  &  00 32 34.81  &  07 29 27.10  &  12.62  &  5.68  &  $>$M5  &  0.925  &  0.008  &  0.045  &  Bc\\
\tiny TYC\,5853\,1318\,1  &  01 07 11.94  &  -19 35 36.00  &  11.41  &  4.16  &  M1V  &  7.26  &  0.07  &  0.10  &  S?\\
\tiny 2MASS J01112542+1526214A  &  01 11 25.42  &  15 26 21.50  &  14.46  &  6.25  &  M5V  &  0.911  &  0.001  &  0.01  &  Bc\\
\tiny 2MASS J01112542+1526214B  &  01 11 25.42  &  15 26 21.50  &  14.46  &  6.55  &  M6V  &  0.791  &  0.001  &  0.01  &  Bc\\
\tiny 2MASS J01132817-3821024  &  01 13 28.17  &  -38 21 02.50  &  11.77  &  4.17  &  (M0V+M3V)+M1V  &  0.446  &  ---  &  0.210  &  Tc\\
\tiny 2MASS J01351393-0712517  &  01 35 13.93  &  -07 12 51.77  &  13.42  &  5.50  &  M4.5V  &  0.703  &  ---  &  0.080  &  SB2\\
\tiny 2MASS J01365516-0647379  &  01 36 55.16  &  -06 47 37.92  &  14.00  &  5.14  &  M4V+L0  &  0.346  &  0.001  &  0.11  &  Bw\\
\tiny TYC\,1208\,0468\,1  &  01 37 39.42  &  18 35 32.91  &  9.83  &  3.11  &  K3V+K5V  &  2.803  &  0.010  &  0.07  &  Bw\\
\tiny 2MASS J01535076-1459503  &  01 53 50.77  &  -14 59 50.30  &  11.97  &  4.90  &  M3V+M3V  &  1.515  &  ---  &  0.110  &  BC\\
\tiny 2MASS J02014677+0117161  &  02 01 46.78  &  01 17 16.20  &  12.78  &  4.51  &  M  & ---  & --- & --- & ---  \\
\tiny RBS\,269  &  02 01 46.93  &  01 17 06.00  &  12.72  &  4.46  &  M  &  5.98/3.30  &  0.01  &  0.09  &  Bw\\
\tiny 2MASS J02175601+1225266  &  02 17 56.01  &  12 25 26.70  &  13.62  &  4.54  &  M3.5V  &  1.995  &  0.005  &  0.05  &  S\\
\tiny HIP\,10679  &  02 17 24.74  &  28 44 30.43  &  7.75  &  1.49  &  G2V  &  0.777  &  0.005  &  0.070  &  Bw+D\\
\tiny HIP\,10680  &  02 17 25.28  &  28 44 42.16  &  6.95  &  1.16  &  F5V  &  0.240  &  0.001  &  0.030  &  Bw\\
\tiny HIP\,11152  &  02 23 26.64  &  +22 44 06.75  &  11.09  &  3.74  &  M3V  &  1.80/3.60  &  0.02  &  0.06  &  S\\
\tiny HIP\,11437A  &  02 27 29.25  &  30 58 24.60  &  10.12  &  3.04  &  K4V  &  12.5  &  0.5  &  0.20  &  Bw+D\\
\tiny HIP\,11437B  &  02 27 28.05  &  30 58 40.53  &  12.44  &  4.22  &  M1V  &  4.66  &  0.05  &  0.16  &  Bw\\
\tiny HIP\,12545  &  02 41 25.90  &  05 59 18.00  &  10.37  &  3.30  &  K6Ve  &  4.83  &  0.03  &  0.180  &  S\\
\tiny 2MASS J03350208+2342356  &  03 35 02.09  &  23 42 35.61  &  17.00  &  5.74  &  M8.5V  &  0.472  &  0.005  &  0.03  &  Bc?\\
\tiny 2MASS J03461399+1709176  &  03 46 14.00  &  17 09 17.45  &  12.90  &  4.08  &  M0.5  &  1.742  &  0.001  &  0.07  &  S\\
\tiny GJ\,3305  &  04 37 37.30  &  -02 29 28.00  &  10.59  &  4.18  &  M1+M?  &  4.89  &  0.01  &  0.05  &  Bc\\
\tiny 2MASS J04435686+3723033  &  04 43 56.87  &  37 23 03.30  &  12.98  &  4.18  &  M3Ve+M5?  &  4.288  &  ---  &  ---  &  Bw\\
\tiny HIP\,23200  &  04 59 34.83  &  01 47 00.68  &  10.05  &  3.99  &  M0.5Ve  &  4.430  &  0.030  &  0.150  &  SB1\\
\tiny TYC\,1281\,1672\,1  &  05 00 49.28  &  15 27 00.71  &  10.75  &  3.15  &  K2IV  &  2.76  &  0.01  &  0.12  &  S\\
\tiny HIP\,23309  &  05 00 47.10  &  -57 15 25.00  &  10.00  &  3.76  &  M0Ve  &  8.60  &  0.07  &  0.110  &  S\\
\tiny 2MASS J05015665+0108429  &  05 01 56.65  &  01 08 42.91  &  13.20  &  5.52  &  M4V  &  2.08  &  0.02  &  0.07  &  S?\\
\tiny HIP\,23418A  &  05 01 58.80  &  09 59 00.00  &  11.45  &  4.78  &  M3V  &  1.220  &  0.010  &  0.070  &  SB2\\
\tiny HIP\,23418B  &  05 01 58.80  &  09 59 00.00  &  12.45  &  5.23  &  $>$M3V  &  ---  &  ---  &  ---  &  Tc\\
\tiny BD\,-21\,1074A  &  05 06 49.90  &  -21 35 09.00  &  10.29  &  4.35  &  M1.5V  &  9.3  &  0.1  &  0.120  &  Tw\\
\tiny BD\,-21\,1074B  &  05 06 49.90  &  -21 35 09.00  &  11.67  &  4.64  &  M2.5V  &  5.40  &  0.10  &  0.080  &  Tc\\
\tiny 2MASS J05082729-2101444  &  05 08 27.30  &  -21 01 44.40  &  14.70  &  5.87  &  M5.6V  &  0.280  &  0.002  &  0.07  &  S\\
\tiny TYC\,1121\,486\,1  &  05 20 31.83  &  +06 16 11.48  &  11.67  &  3.11  &  K4V  &  2.18  &  ---  &  0.09  &  Tc\\
\tiny TYC\,112\,917\,1  &  05 20 00.29  &  +06 13 03.57  &  11.58  &  3.00  &  K4V  &  3.51  &  ---  &  0.08  &  Tw\\
\tiny 2MASS J05241914-1601153  &  05 24 19.15  &  -16 01 15.30  &  14.32  &  5.60  &  M4.5+M5  &  0.401  &  0.001  &  0.15  &  Bc\\
\tiny HIP\,25486  &  05 27 04.76  &  -11 54 03.47  &  6.22  &  1.29  &  F7V  &  0.966  &  0.002  &  0.10  &  SB2\\
\tiny 2MASS J05294468-3239141  &  05 29 44.68  &  -32 39 14.20  &  13.79  &  5.47  &  M4.5V  &  1.532  &  0.005  &  0.03  &  S?\\
\tiny TYC\,4770\,0797\,1  &  05 32 04.51  &  -03 05 29.38  &  11.32  &  4.31  &  M2V+M3.5V  &  4.372  &  0.002  &  0.160  &  Bc\\
\tiny 2MASS J05335981-0221325  &  05 33 59.81  &  -02 21 32.50  &  12.42  &  4.72  &  M2.9V  &  7.250  &  ---  &  0.170  &  S\\
\tiny 2MASS J06131330-2742054  &  06 13 13.31  &  -27 42 05.50  &  12.09  &  5.23  &  M3.V:  &  16.8  &  1.0  &  0.07  &  Tc\\
\tiny HIP\,29964  &  06 18 28.20  &  -72 02 41.00  &  9.80  &  2.99  &  K4Ve  &  2.670  &  0.010  &  0.120  &  S+D\\
\tiny 2MASS J07293108+3556003AB  &  07 29 31.09  &  35 56 00.40  &  11.82  &  4.02  &  M1+M3  &  1.970  &  0.010  &  0.10  &  Bc\\
\tiny 2MASS J08173943-8243298  &  08 17 39.44  &  -82 43 29.80  &  11.62  &  5.03  &  M3.5V  &  1.318  &  ---  &  0.050  &  Bc?\\
\tiny 2MASS J08224744-5726530  &  08 22 47.45  &  -57 26 53.00  &  13.37  &  5.57  &  M4.5+L0  &  ---  &  ---  &  ---  &  Tc\\
\tiny 2MASS J09361593+3731456AB  &  09 36 15.91  &  37 31 45.50  &  11.09  &  4.10  &  M0.5+M0.5  &  12.9  &  0.3  &  0.030  &  SB2\\
\tiny 2MASS J10015995+6651278  &  10 02 00.10  &  66 51 26.00  &  12.38  &  4.16  &  M3  &  2.49  &  0.02  &  0.060  &  Bc?\\
\tiny HIP\,50156  &  10 14 19.17  &  21 04 29.55  &  10.08  &  3.82  &  M0.5V+?  &  7.860  &  ---  &  0.050  &  Bc\\
\tiny TWA\,22  &  10 17 26.89  &  -53 54 26.50  &  13.99  &  6.30  &  M5+M6  &  0.830  &  0.010  &  0.020  &  Bc\\
\tiny BD\,+26\,2161A  &  10 59 38.31  &  25 26 15.50  &  8.45  &  2.61  &  K2  &  2.022/0.974  &  0.005  &  0.010  &  Bw+D\\
\tiny BD\,+26\,2161B  &  10 59 38.31  &  25 26 15.50  &  9.09  &  3.25  &  K5  &  0.974/2.022  &  0.005  &  0.010  &  Bw\\
\tiny 2MASS J11515681+0731262  &  11 51 56.81  &  07 31 26.25  &  12.38  &  4.61  &  M2+M2+M8  &  2.291  &  ---  &  0.130  &  SB2\\
\tiny 2MASS J13545390-7121476  &  13 54 53.90  &  -71 21 47.67  &  12.24  &  4.57  &  M2.5V  &  3.65  &  0.02  &  0.020  &  S?\\
\tiny HIP\,69562A  &  14 14 21.36  &  -15 21 21.75  &  10.27  &  3.67  &  K5.5V+  &  0.298  &  0.005  &  0.17  &  Tc\\
\tiny HIP\,69562B  &  14 14 21.36  &  -15 21 21.75  &  10.27  &  3.67  &  ---  &  ---  &  ---  &  ---  &  Tc\\
\tiny TYC\,915\,1391\,1  &  14 25 55.93  &  14 12 10.14  &  10.89  &  3.60  &  K4V  &  4.340  &  ---  &  0.360  &  S\\
\tiny HIP\,76629  &  15 38 57.50  &  -57 42 27.00  &  7.97  &  2.12  &  K0V  &  4.27  &  0.10  &  0.180  &  SB1\\
\tiny 2MASS J16430128-1754274  &  16 43 01.29  &  -17 54 27.50  &  12.50  &  3.95  &  M0.6  &  5.14  &  0.04  &  0.140  &  S\\
\tiny 2MASS J16572029-5343316  &  16 57 20.30  &  -53 43 31.70  &  12.44  &  4.65  &  M3V  &  7.15  &  0.05  &  0.020  &  S\\
\tiny 2MASS J17150219-3333398  &  17 15 02.20  &  -33 33 39.80  &  10.93  &  3.86  &  M0V  &  0.311  &  ---  &  0.110  &  Bc?\\
\tiny HIP\,84586  &  17 17 25.50  &  -66 57 04.00  &  7.23  &  2.53  &  G5IV+K5IV  &  1.680  &  0.010  &  0.120  &  SB2\\
\tiny HD\,155555C  &  17 17 31.29  &  -66 57 05.49  &  12.71  &  5.08  &  M3.5Ve  &  4.43  &  0.01  &  0.070  &  Tw\\
\tiny TYC\,8728\,2262\,1  &  17 29 55.10  &  -54 15 49.00  &  9.55  &  2.19  &  K1V  &  1.775  &  0.005  &  0.150  &  S\\
\tiny GSC\,08350-01924  &  17 29 20.67  &  -50 14 53.00  &  13.47  &  4.77  &  M3V  &  1.906  &  0.005  &  0.05  &  Bc\\
\tiny HD\,160305  &  17 41 49.03  &  -50 43 28.00  &  8.35  &  1.36  &  F9V  &  1.336  &  0.008  &  0.060  &  S+D\\
\tiny TYC\,8742\,2065\,1  &  17 48 33.70  &  -53 06 43.00  &  8.94  &  2.16  &  K0IV+  &  2.60/1.62  &  0.01  &  0.060  &  Tc\\
\tiny HIP\,88399  &  18 03 03.41  &  -51 38 56.43  &  12.50  &  4.23  &  M2V+F6V  &  ---  &  ---  &  ---  &  Bw\\
\tiny V4046\,Sgr  &  18 14 10.50  &  -32 47 33.00  &  10.44  &  3.19  &  K5V+K7V  &  2.42  &  0.01  &  0.090  &  SB2+D\\
\tiny UCAC2\,18035440  &  18 14 22.07  &  -32 46 10.12  &  12.78  &  4.24  &  M1Ve  &  12.05  &  0.5  &  0.14  &  SB\\
\tiny 2MASS J18151564-4927472  &  18 15 15.64  &  -49 27 47.20  &  12.86  &  4.78  &  M3V  &  0.447  &  0.002  &  0.130  &  SB1\\
\tiny HIP\,89829  &  18 19 52.20  &  -29 16 33.00  &  8.89  &  1.84  &  G1V  &  0.571  &  0.001  &  0.140  &  S\\
\tiny 2MASS J18202275-1011131A  &  18 20 22.74  &  -10 11 13.62  &  10.63  &  3.35  &  K5Ve  &  4.65/5.15  &  ---  &  0.070  &  Bw+D\\
\tiny 2MASS J18202275-1011131B  &  18 20 22.74  &  -10 11 13.62  &  10.63  &  4.01  &  K7Ve  &  5.15/4.65  &  ---  &  0.070  &  Bw\\
\tiny 2MASS J18420694-5554254  &  18 42 06.95  &  -55 54 25.50  &  13.53  &  4.95  &  M3.5V  &  5.403  &  ---  &  0.070  &  S?\\
\tiny TYC\,9077\,2489\,1  &  18 45 37.02  &  -64 51 46.14  &  9.30  &  3.20  &  K8Ve  &  0.345  &  0.005  &  0.160  &  Tc\\
\tiny TYC\,9073\,0762\,1  &  18 46 52.60  &  -62 10 36.00  &  11.80  &  3.95  &  M1Ve  &  5.37  &  0.04  &  0.320  &  S\\
\tiny HD\,173167  &  18 48 06.36  &  -62 13 47.02  &  7.28  &  1.14  &  F5V  &  0.290  &  0.005  &  0.220  &  SB1\\
\tiny TYC\,740800541  &  18 50 44.50  &  -31 47 47.00  &  11.20  &  3.66  &  K8Ve  &  1.075  &  0.005  &  0.150  &  EB\\
\tiny HIP\,92680  &  18 53 05.90  &  -50 10 50.00  &  8.29  &  1.92  &  K8Ve  &  0.944  &  0.001  &  0.110  &  Bw\\
\tiny TYC\,6872\,1011\,1  &  18 58 04.20  &  -29 53 05.00  &  11.78  &  3.76  &  M0Ve  &  0.503  &  0.004  &  0.060  &  Bw\\
\tiny 2MASS J19102820-2319486  &  19 10 28.21  &  -23 19 48.60  &  13.20  &  4.99  &  M4V  &  3.64  &  0.02  &  0.13  &  S\\
\tiny TYC\,6878\,0195\,1  &  19 11 44.70  &  -26 04 09.00  &  10.27  &  2.90  &  K4Ve  &  5.70  &  0.05  &  0.090  &  Bw\\
\tiny 2MASS J19233820-4606316  &  19 23 38.20  &  -46 06 31.60  &  11.87  &  3.60  &  M0V  &  3.237  &  ---  &  0.110  &  S\\
\tiny 2MASS J19243494-3442392  &  19 24 34.95  &  -34 42 39.30  &  14.28  &  5.50  &  M4V  &  0.708  &  0.001  &  0.020  &  Bc?\\
\tiny TYC\,7443\,1102\,1  &  19 56 04.37  &  -32 07 37.71  &  11.80  &  3.95  &  M0.0V  &  11.3  &  0.2  &  0.09  &  Tw\\
\tiny 2MASS J19560294-3207186AB  &  19 56 02.94  &  -32 07 18.70  &  13.30  &  5.12  &  M4V  &  1.569  &  0.003  &  0.030  &  Tc\\
\tiny 2MASS J20013718-3313139  &  20 01 37.18  &  -33 13 14.01  &  12.25  &  4.06  &  M1V  &  12.7  &  0.2  &  0.13  &  Tw\\
\tiny 2MASS J20055640-3216591  &  20 05 56.41  &  -32 16 59.15  &  11.96  &  4.02  &  M2V  &  8.368  &  0.005  &  0.130  &  S\\
\tiny HD\,191089  &  20 09 05.21  &  -26 13 26.52  &  7.18  &  1.10  &  F5V  &  0.488  &  0.005  &  ---  &  S+D\\
\tiny 2MASS J20100002-2801410AB  &  20 10 00.03  &  -28 01 41.10  &  13.62  &  4.64  &  M2.5+M3.5  &  0.470  &  0.005  &  0.040  &  Bc\\
\tiny 2MASS J20333759-2556521  &  20 33 37.59  &  -25 56 52.20  &  14.87  &  5.99  &  M4.5V  &  0.710  &  0.001  &  0.05  &  S\\
\tiny HIP\,102141A  &  20 41 51.20  &  -32 26 07.00  &  11.09  &  5.42  &  M4Ve  &  1.191  &  0.005  &  0.040  &  Bc\\
\tiny HIP\,102141B  &  20 41 51.10  &  -32 26 10.00  &  11.13  &  5.42  &  M4Ve  &  0.781  &  0.002  &  0.020  &  Bc\\
\tiny 2MASS J20434114-2433534  &  20 43 41.14  &  -24 33 53.19  &  12.83  &  4.97  &  M3.7+M4.1  &  1.610  &  0.010  &  0.03  &  Bc\\
\tiny HIP\,102409  &  20 45 09.50  &  -31 20 27.00  &  8.73  &  4.20  &  M1Ve  &  4.86  &  0.02  &  0.10  &  S+D\\
\tiny HIP\,103311  &  20 55 47.67  &  -17 06 51.04  &  7.35  &  1.54  &  F8V  &  0.356  &  0.004  &  0.06  &  Bc\\
\tiny TYC\,6349\,0200\,1  &  20 56 02.70  &  -17 10 54.00  &  10.62  &  3.54  &  K6Ve+M2  &  3.41  &  0.05  &  0.120  &  Bw\\
\tiny 2MASS J21100535-1919573  &  21 10 05.36  &  -19 19 57.40  &  11.54  &  4.34  &  M2V  &  3.71  &  0.02  &  0.29  &  S\\
\tiny 2MASS J21103147-2710578  &  21 10 31.48  &  -27 10 57.80  &  15.20  &  5.60  &  M4.5V  &  1.867  &  0.008  &  0.04  &  Bw\\
\tiny 2MASS J21103096-2710513  &  21 10 30.96  &  -27 10 51.30  &  15.72  &  5.60  &  M5V  &  ---  &  ---  &  ---  &  Bw\\
\tiny HIP\,105441  &  21 21 24.49  &  -66 54 57.37  &  8.77  &  2.37  &  K2V  &  5.50  &  0.02  &  0.050  &  Bw\\
\tiny TYC\,9114\,1267\,1  &  21 21 28.72  &  -66 55 06.30  &  10.59  &  3.58  &  K7V  &  20.5  &  1.0  &  0.015  &  Bw\\
\tiny TYC\,9486\,927\,1  &  21 25 27.49  &  -81 38 27.68  &  11.70  &  4.36  &  M1V  &  0.542  &  ---  &  0.190  &  Bc\\
\tiny 2MASS J21374019+0137137AB  &  21 37 40.19  &  01 37 13.70  &  13.36  &  5.48  &  M5V  &  0.202  &  0.001  &  0.130  &  Bc\\
\tiny 2MASS J21412662+2043107  &  21 41 26.63  &  20 43 10.70  &  13.50  &  4.89  &  M3V  &  0.899  &  0.001  &  0.03  &  Bc?\\
\tiny TYC\,2211\,1309\,1  &  22 00 41.59  &  27 15 13.60  &  11.39  &  3.67  &  M0V  &  1.109  &  0.001  &  0.080  &  Bc\\
\tiny TYC\,9340\,0437\,1  &  22 42 48.90  &  -71 42 21.00  &  10.60  &  3.71  &  K7Ve  &  4.46  &  0.03  &  0.16  &  S\\
\tiny HIP\,112312  &  22 44 58.00  &  -33 15 02.00  &  12.10  &  5.17  &  M4Ve  &  2.37  &  0.01  &  0.110  &  Bw\\
\tiny TX\,Psa  &  22 45 00.05  &  -33 15 25.80  &  13.36  &  5.57  &  M4.5Ve  &  1.080  &  0.005  &  0.030  &  Bw\\
\tiny 2MASS J22571130+3639451  &  22 57 11.31  &  36 39 45.14  &  12.50  &  3.86  &  M3V  &  1.220  &  0.020  &  0.04  &  S\\
\tiny TYC\,5832\,0666\,1  &  23 32 30.90  &  -12 15 52.00  &  10.54  &  3.97  &  M0Ve  &  5.68  &  0.03  &  0.140  &  S\\
\tiny 2MASS\,J23500639+2659519  &  23 50 06.39  &  26 59 51.93  &  14.26  &  4.96  &  M3.5V  &  0.287  &  0.005  &  0.05  &  Bc?\\
\tiny 2MASS J23512227+2344207  &  23 51 22.28  &  23 44 20.80  &  14.11  &  5.29  &  M4V  &  3.208  &  0.004  &  0.060  &  S\\
\enddata
\end{deluxetable}
\clearpage
\newpage
\setcounter{table}{3}
\begin{deluxetable}{ccccccccccc}
\tablecolumns{11}
\tablecaption{\label{tab_period2} Results of membership assessment based on velocity (U, V, W), space (X, Y, Z) components, Li EW, and rotation period (P)
}
\tablewidth{0pt}
 \tablehead{\colhead{Target} & \colhead{U} & \colhead{V} & \colhead{W} &  \colhead{X} & \colhead{Y} & \colhead{Z} & \colhead{Li} & \colhead{P}  & \colhead{final} & \colhead{Note} \\
}
\startdata
  \hline

HIP\,560 & Y & Y & Y & Y & Y & Y & Y & Y & Y & Core \\
2MASS\,J00172353-6645124 & Y & Y & Y & Y & Y & Y & - & Y & Y & Core \\
TYC\,1186\,0706\,1 & Y & Y & Y & Y & Y & Y & Y & Y & Y  & \\
GJ\,2006A & Y & N & Y & Y & Y & Y & Y & Y & Y & Core \\
GJ\,2006B & Y & Y & Y & Y & Y & Y & Y & Y & Y & Core \\
2MASS\,J00323480+072927A & Y & Y & Y & Y & Y & Y & - & ? & N &  \\
2MASS\,J00323480+072927B & Y & Y & Y & Y & Y & Y & - & ? & N &  \\
TYC\,5853\,1318\,1 & N & N & Y & Y & Y & Y & - & Y & Y & C  \\
2MASS\,J01112542+1526214A & Y & Y & Y & Y & Y & Y & Y & - & Y &  \\
2MASS\,J01112542+1526214B & Y & Y & Y & Y & Y & Y & - & - & Y &  \\
2MASS\,J01132817-3821024 & Y & Y & Y & Y & Y & Y & - & Y & Y &  \\
2MASS\,J01351393-0712517 & Y & Y & Y & Y & Y & Y & Y & - & Y  & \\
2MASS\,J01365516-0647379 & Y & Y & Y & Y & Y & Y & N & N & N &  \\
TYC\,1208\,0468\,1 & Y & Y & Y & Y & Y & Y & Y & N & Y & C  \\
2MASS\,J01535076-1459503 & Y & Y & Y & Y & Y & Y & - & Y & Y  & \\
2MASS\,J02014677+0117161 & Y & N & Y & Y & Y & N & - & Y & Y & Core\_e \\
RBS\,269 & Y & N & Y & Y & Y & N & - & Y & Y& Core\_e \\
2MASS\,J02175601+1225266 & Y & N & Y & Y & Y & N & - & N & Y & Core\_e \\
HIP\,10679 & Y & Y & Y & Y & Y & Y & Y & Y & Y & Core \\
HIP\,10680 & Y & Y & Y & Y & Y & Y & Y & Y & Y & Core \\
HIP\,11152 & Y & Y & Y & Y & Y & Y & - & Y & Y & Core \\
HIP\,11437A & Y & Y & Y & Y & Y & Y & N & N & Y & Core \\
HIP\,11437B & Y & Y & Y & Y & Y & Y & Y & Y & Y & Core \\
HIP\,12545 & Y & Y & Y & Y & Y & Y & Y & Y & Y & Core \\
2MASS\,J03350208+2342356 & Y & N & Y & Y & Y & Y & - & Y & Y & C \\
2MASS\,J03461399+1709176 & N & N & N & Y & Y & Y & - & N & N &  \\
GJ\,3305 & N & Y & Y & Y & Y & Y & Y & U & ? & C \\
2MASS\,J04435686+3723033 & Y & N & N & Y & Y & Y & Y & Y & Y & Core\_e \\
HIP\,23200 & Y & Y & Y & Y & Y & Y & Y & - & Y &  \\
TYC\,1281\,1672\,1 & Y & Y & Y & Y & Y & Y & - & Y & Y &  \\
HIP\,23309 & Y & Y & Y & Y & Y & Y & Y & Y & Y &  \\
2MASS\,J05015665+0108429 & Y & Y & Y & Y & Y & Y & - & Y & Y &  \\
HIP\,23418A & Y & Y & Y & Y & Y & Y & Y & - & Y &  \\
HIP\,23418B & Y & Y & Y & Y & Y & Y & - & - & Y &  \\
BD\,-21\,1074A & Y & Y & Y & Y & Y & Y & Y & N & Y & Core \\
BD\,-21\,1074B & Y & Y & N & Y & Y & Y & Y & N & N & C  \\
2MASS\,J05082729-2101444 & Y & Y & Y & Y & Y & Y & Y & Y & Y & Core \\
TYC\,112\,1486\,1 & Y & Y & Y & Y & Y & Y & - & Y & Y &  \\
TYC\,112\,917\,1 & Y & Y & Y & Y & Y & Y & - & Y & Y &  \\
2MASS\,J05241914-1601153 & Y & Y & Y & Y & Y & Y & Y & Y & Y &  \\
HIP\,25486 & Y & Y & Y & Y & Y & Y & Y & - & Y &  \\
2MASS\,J05294468-3239141 & N & N & N & Y & Y & Y & - & Y & Y &  C \\
TYC\,4770\,0797\,1 & Y & Y & Y & Y & Y & Y & - & N & N &  C \\
2MASS\,J05335981-0221325 & Y & Y & Y & Y & Y & Y & Y & Y & Y & Core \\
2MASS\,J06131330-2742054 & Y & Y & Y & Y & Y & Y & Y & N & Y &  \\
HIP\,29964 & Y & Y & Y & Y & Y & Y & Y & Y & Y & Core \\
2MASS\,J07293108+3556003AB & Y & Y & Y & Y & Y & N & - & Y & Y & C \\
2MASS\,J08173943-8243298 & Y & Y & Y & Y & Y & Y & - & Y & Y &  \\
2MASS\,J08224744-5726530 & Y & Y & Y & Y & Y & Y & - & - & Y &  \\
2MASS\,J09361593+3731456AB & Y & Y & N & Y & Y & N & - & - & - & C \\
2MASS\,J10015995+6651278 & Y & N & Y & Y & Y & N & - & Y & Y & C \\
HIP\,50156 & Y & N & N & Y & Y & N & - & N & N & C \\
TWA\,22 & Y & Y & Y & Y & Y & Y & Y & - & Y &  \\
BD\,+26\,2161A & N & N & N & N & N & N & - & Y & N &  \\
BD\,+26\,2161B & Y & Y & Y & Y & Y & N & - & N & N & C  \\
2MASS\,J11515681+0731262 & Y & Y & N & Y & Y & N & - & - &  & C \\
2MASS\,J13545390-7121476 & Y & Y & Y & Y & Y & Y & - & Y & Y &  \\
HIP\,69562A & Y & N & N & Y & Y & N & - & Y & Y & C \\
HIP\,69562B & Y & N & Y & Y & Y & N & - & - & Y &  C \\
TYC\,915\,1391\,1 & N & N & N & Y & Y & N & Y & Y & N &  \\
HIP\,76629 & Y & Y & Y & Y & Y & Y & - & - & Y &  \\
2MASS\,J16430128-1754274 & Y & N & N & Y & Y & N & Y & Y & Y & Core\_e \\
2MASS\,J16572029-5343316 & Y & Y & Y & Y & Y & Y & - & Y & Y & C \\
2MASS\,J17150219-3333398 & Y & Y & Y & Y & Y & Y & - & Y & Y &  \\
HIP\,84586 & Y & Y & Y & Y & Y & Y & N & - & Y &  \\
HD\,155555C & Y & Y & Y & Y & Y & Y & Y & Y & Y & Core \\
TYC\,8728\,2262\,1 & Y & Y & Y & Y & Y & Y & Y & Y & Y &  \\
GSC\,08350-01924 & Y & Y & Y & Y & Y & Y & Y & Y & Y &  \\
HD\,160305 & Y & Y & Y & Y & Y & Y & - & N & Y &  \\
TYC\,8742\,2065\,1 & Y & Y & Y & Y & Y & Y & - & N & Y &  \\
HIP\,88399 & Y & Y & Y & Y & Y & Y & - & Y & Y & Core \\
V4046\,Sgr & Y & Y & Y & Y & Y & Y & Y & - & Y &  \\
UCAC2\,18035440 & N & N & N & N & N & N & - & - & N &  \\
2MASS\,J18151564-4927472 & N & Y & Y & Y & Y & Y & Y & - & - & C  \\
HIP\,89829 & Y & Y & Y & Y & Y & Y & Y & Y & Y & Core \\
2MASS\,J18202275-1011131A & N & Y & Y & Y & Y & N & Y & Y & Y & C  \\
2MASS\,J18202275-1011131B & N & Y & Y & Y & Y & N & - & Y & Y &  C \\
2MASS\,J18420694-5554254 & Y & Y & Y & Y & Y & Y & - & Y & Y &  \\
TYC907724891 & Y & Y & Y & Y & Y & Y & Y & Y & Y &  \\
TYC\,9073\,0762\,1 & Y & Y & Y & Y & Y & Y & Y & Y & Y & Core \\
HD\,173167 & Y & Y & Y & Y & Y & Y & Y & - & Y &  \\
TYC\,7408\,0054\,1 & Y & Y & Y & Y & Y & Y & Y & - & Y &  \\
HIP\,92680 & Y & Y & Y & Y & Y & Y & Y & Y & Y & Core \\
TYC\,6872\,1011\,1 & N & Y & Y & Y & Y & Y & Y & N & Y & Core\_e \\
2MASS\,J19102820-2319486 & Y & Y & Y & Y & Y & Y & Y & Y & Y & Core \\
TYC\,6878\,0195\,1 & Y & Y & Y & Y & Y & Y & N & N & Y & Core \\
2MASS\,J19233820-4606316 & Y & Y & Y & Y & Y & Y & Y & Y & Y & Core \\
2MASS\,J19243494-3442392 & Y & Y & Y & Y & Y & Y & - & Y & Y &  \\
TYC\,7443\,1102\,1 & Y & Y & Y & Y & Y & Y & Y & N & Y & Core \\
2MASS\,J19560294-3207186AB & Y & Y & Y & Y & Y & Y & Y & Y & Y &  \\
2MASS\,J20013718-3313139 & Y & Y & Y & Y & Y & Y & Y & N & Y & Core \\
2MASS\,J20055640-3216591 & N & N & N & N & N & N & Y & Y & N &  \\
HD\,191089 & Y & Y & Y & Y & Y & Y & Y & Y & Y & Core \\
2MASS\,J20100002-2801410AB & Y & Y & Y & Y & Y & Y & Y & Y & Y &  \\
2MASS\,J20333759-2556521 & Y & Y & Y & Y & Y & Y & Y & Y & Y & Core \\
HIP\,102141A & Y & Y & Y & Y & Y & Y & Y & Y & Y &  \\
HIP\,102141B & Y & Y & Y & Y & Y & Y & Y & Y & Y &  \\
2MASS\,J20434114-2433534 & Y & N & N & Y & Y & Y & Y & Y & Y & C \\
HIP\,102409 & Y & Y & Y & Y & Y & Y & Y & Y & Y & Core \\
HIP\,103311 & N & Y & N & Y & Y & Y & Y & Y & Y & C \\
TYC\,6349\,0200\,1 & Y & Y & Y & Y & Y & Y & Y & Y & Y & Core \\
2MASS\,J21100535-1919573 & Y & Y & Y & Y & Y & Y & Y & Y & Y & Core \\
2MASS\,J21103147-2710578 & Y & Y & Y & Y & Y & Y & Y & Y & Y & Core \\
2MASS\,J21103096-2710513 & Y & Y & Y & Y & Y & Y & - & Y & Y & Core \\
HIP\,105441 & Y & Y & Y & Y & Y & Y & N & N & N &  \\
TYC\,9114\,1267\,1 & Y & Y & Y & Y & Y & Y & N & N & N &  \\
TYC\,9486\,927\,1 & N & Y & Y & Y & Y & Y & Y & Y & Y & C  \\
2MASS\,J21374019+0137137AB & Y & N & N & Y & Y & Y & - & Y & Y & C \\
2MASS\,J21412662+2043107 & Y & Y & Y & Y & N & Y & - & Y & Y & C \\
TYC\,2211\,1309\,1 & Y & N & Y & Y & Y & Y & Y & - &  - &  C  \\
TYC\,9340\,0437\,1 & Y & Y & Y & Y & Y & Y & Y & Y & Y & Core \\
HIP\,112312 & Y & Y & Y & Y & Y & Y & Y & Y & Y & Core \\
TX\,Psa & Y & Y & Y & Y & Y & Y & Y & N & Y & Core \\
2MASS\,J22571130+3639451 & Y & N & N & Y & N & Y & - & N & N &  \\
TYC\,5832\,0666\,1 & Y & Y & Y & Y & Y & Y & Y & Y & Y & Core \\
2MASS\,J23500639+2659519 & Y & Y & Y & Y & Y & Y & - & Y & Y &  \\
2MASS\,J23512227+2344207 & Y & Y & Y & Y & Y & Y & - & Y & Y &  C \\
\hline
\multicolumn{10}{l}{Core: core member; Core\_e: core member excluded from fit; C: candidate}\\
\enddata
\end{deluxetable}

\end{document}